\documentclass[journal]{IEEEtran}
\usepackage[style=ieee,sorting=none,maxnames=4,minnames=3,maxbibnames=3]{biblatex}

\usepackage{amsfonts}
\usepackage[cmex10]{amsmath}
\usepackage{amsthm}

\theoremstyle{definition}

\newtheorem{assumption}{Assumption}
\newtheorem{remark}{Remark}
\newtheorem{theorem}{Theorem}
\newtheorem{corollary}{Corollary}

\theoremstyle{remark}

\usepackage{braket}
\usepackage{array}
\usepackage{mdwmath}
\usepackage{mathrsfs}
\usepackage{mdwtab}
\usepackage{eqparbox}
\usepackage[caption=1]{caption}
\usepackage{stfloats}
\usepackage{url}
\usepackage{amssymb}
\usepackage{float}
\usepackage{indentfirst}
\usepackage{makeidx}
\usepackage{tabularx}
\usepackage{cases}
\usepackage{bbm}
\usepackage{algorithmic}
\usepackage{mathtools}
\usepackage{color}
\usepackage{bm}
\usepackage{stfloats}
\usepackage{lipsum}
\usepackage{amsfonts}
\usepackage{bm}
\usepackage{dsfont}
\usepackage{lipsum}
\usepackage{graphicx}
\usepackage{url}
\usepackage{bbm}
\newcommand{\RNum}[1]{\uppercase\expandafter{\romannumeral #1\relax}}
\ifCLASSOPTIONcompsoc
\usepackage[caption=false, font=normalsize, labelfont=sf, textfont=sf]{subfig}
\else
\usepackage[caption=false, font=footnotesize]{subfig}
\fi
\usepackage{multirow}
\usepackage{booktabs}
\allowdisplaybreaks
\makeatletter

\usepackage[linesnumbered,ruled,vlined]{algorithm2e}
\usepackage{amsmath}
\makeatother
\usepackage{amsthm}
\theoremstyle{definition} 
\theoremstyle{remark} 
\newif\ifgthshowchanges
\gthshowchangesfalse
\definecolor{gthrevision}{RGB}{0,80,180}
\newcommand{\gthchanged}[1]{{\ifgthshowchanges\color{gthrevision}\fi#1}}
\newcommand{\gthchangebegin}{\begingroup\ifgthshowchanges\color{gthrevision}\fi}
\newcommand{\gthchangeend}{\endgroup}
\usepackage[stretch=10,shrink=30,step=1]{microtype}
\begin{document}
\title{
\gthchanged{Leveraging Quantum Measurement Randomness for Shot-Efficient Diffusion Models}
}

\author{Tianhao Guo,~\IEEEmembership{Member,~IEEE,}
Dianxin Zheng, Yuhua Qian, Yaohui Zheng,
Yuzhi Yang*,~\IEEEmembership{Member,~IEEE,}
M\'erouane Debbah,~\IEEEmembership{Fellow,~IEEE,}
and Chau Yuen,~\IEEEmembership{Fellow,~IEEE}
\vspace{-0.8cm}
\thanks{
Tianhao Guo and Dianxin Zheng are with the College of Physics and Electronic Engineering
and the Shanxi Key Laboratory of Wireless Communication and Detection,
Shanxi University, Taiyuan 030006, China
(e-mails: tianhao\_guo@sxu.edu.cn; 511207402@qq.com).
}
\thanks{
Tianhao Guo is also with the National Key Laboratory of Integrated Services Networks,
Xidian University, Xi'an 710071, China.
}
\thanks{
Yuhua Qian is with the Institute of Big Data Science and Industry,
Key Laboratory of Evolutionary Science Intelligence of Shanxi Province,
School of Artificial Intelligence,
Shanxi University, Taiyuan 030006, China
(e-mail: jinchengqyh@126.com).
}
\thanks{
Yaohui Zheng is with the State Key Laboratory of Quantum Optics Technologies and Devices,
Institute of Opto-Electronics,
Shanxi University, Taiyuan 030006, China
(e-mail: yhzheng@sxu.edu.cn).
}
\thanks{
Yuzhi Yang and M\'erouane Debbah are with the Institute of Digital Future, Khalifa University, Abu Dhabi, UAE
(e-mails: yangyuzhi.zju@gmail.com; merouane.debbah@ku.ac.ae).
(Corresponding author: Yuzhi Yang.)
}
\thanks{
Chau Yuen is with the School of Electrical and Electronics Engineering,
Nanyang Technological University, Singapore 639798
(e-mail: chau.yuen@ntu.edu.sg).
}
}

\maketitle
\setlength{\abovedisplayskip}{5pt plus 2pt minus 1pt}
\setlength{\belowdisplayskip}{5pt plus 2pt minus 1pt}
\setlength{\abovedisplayshortskip}{0pt plus 2pt}
\setlength{\belowdisplayshortskip}{4pt plus 2pt minus 1pt}
\gthchangebegin
\begin{abstract}
In quantum computing, the readout always carries finite-shot sampling noise. Suppressing this noise through repeated measurements raises the cost of a quantum neural network (QNN). Meanwhile, a stochastic diffusion sampler deliberately adds randomness at each reverse (generation) step. This paper turns the measurement noise from a liability into a resource. Specifically, we use the noise quantum measurement produces anyway to replace part of the noise the sampler would otherwise inject. We characterize the measurement error as multinomial sampling noise propagated through a trainable linear readout.
To connect the circuit's Born probabilities to the denoising task, we further formulate the diffusion process on a bounded zero-sum hyperplane mapped to the probability simplex. Therefore, we can employ a compensation noise to match the mean and covariance of the shot-noise-free linear update under some minimum shot conditions. Solving this condition for the shot count gives a state-dependent adaptive budget. For the denoising diffusion implicit model (DDIM), this budget increases as the sampler becomes more deterministic. On an eight-dimensional Gaussian mixture, the schedule achieves comparable generation quality with $303$ shots per sample against a high-shot reference of $1.6\times10^{6}$, about $5300\times$ fewer. Latent diffusion experiments on MNIST further validate the proposed mechanism for generation in a learned latent space.
\end{abstract}
\gthchangeend
\vspace{-0.25cm}
\section{Introduction}
\subsection{Motivation}
Quantum computing (QC) has attracted considerable attention due to its potential in computational advantages for certain classes of problems \cite{nielsen2010,preskill2018}. As a framework that combines parameterized quantum circuits (PQCs) with modern machine-learning methods, quantum neural networks (QNNs) have consequently become an important research direction. Among their applications, quantum generative modeling aims to exploit quantum states, circuit dynamics, and measurement statistics to learn complex data distributions.

\gthchanged{The output of a quantum circuit is generally not available as an exact real-valued vector. Instead, the circuit must be measured repeatedly, and its output probabilities are estimated from a finite number of measurement shots. The resulting estimator inevitably contains finite-shot fluctuations. Taking more shots reduces these fluctuations but raises the cost of each QNN evaluation. Therefore, readout precision must be paid for in repeated circuit executions.}

\gthchanged{Meanwhile, diffusion models (DMs) have achieved remarkable success in learning high-dimensional data distributions \cite{sohldickstein2015,ddpm,song2021sde,rombach2022ldm}. Unlike one-step direct generation methods, a denoising DM generates a sample through a sequence of reverse denoising steps rather than a single forward evaluation. Starting from random noise, the sampler repeatedly uses a denoiser to predict a clean sample from the current state and diffusion time. In this work, a QNN supplies this per-step denoising prediction, while the complete reverse loop generates a new sample. Stochastic samplers also inject noise during the reverse process, which provides the budget we seek to reuse.}

\gthchanged{Stochastic diffusion sampling introduces a tension into this precision requirement. A conventional QNN-based sampler first spends measurements suppressing fluctuations in the denoising prediction, then explicitly injects fresh random noise into the reverse update. Some of the randomness removed at the readout is thus followed immediately by randomness added by the sampler. This raises the question motivating our work: can finite-shot measurement fluctuations supply part of the prescribed reverse-step noise, allowing us to measure less without sacrificing generation quality?}

\gthchanged{A second issue is the mismatch between quantum measurement outputs and the data representation used by DMs. Standard DMs operate in Euclidean spaces, where coordinates may take positive or negative values and need not sum to one. In contrast, computational-basis measurements of a quantum circuit naturally produce probability vectors on the simplex. Although amplitude encoding can represent signed information at the state-vector level, direct measurements still return probabilities. Angular encoding is straightforward to implement using parameterized rotations, but generally requires a number of qubits that grows linearly with the input dimension. Probability encoding is therefore attractive for direct quantum readout, provided that Euclidean data and the denoising target can be connected to it.}

\gthchanged{To reconcile these representations, we formulate the diffusion process on a bounded zero-sum hyperplane and connect this space to the probability simplex through an affine transformation. Both the noisy input and the clean denoising target can then be represented as probability vectors, without reconstructing signed amplitudes from measurement outcomes. A trainable linear readout maps the circuit probabilities back into the data space. Finite-shot readout is characterized by multinomial sampling, so its covariance can be propagated explicitly through this map and compared with the noise prescribed by the reverse step.}

The central idea of this work is to align this finite-shot measurement noise with the stochasticity of the reverse diffusion process. When the QNN output is estimated using a limited number of shots, the estimator contains a random perturbation whose covariance is determined by the predicted probability distribution and the measurement budget. By accounting for this covariance in the reverse update, the measurement-induced perturbation can contribute to the total stochasticity required by the sampler, and thereby replace part of the externally injected diffusion noise. Under-sampling is thus no longer viewed merely as an approximation to exact quantum readout, but as a mechanism for redistributing the required randomness between quantum measurement and classical noise injection.

This perspective naturally leads to a shot-adaptive quantum diffusion sampler. In reverse steps with a relatively large stochastic noise budget, fewer measurements may be used because a larger portion of the finite-shot uncertainty can be absorbed into the sampling dynamics. In steps that require a more accurate denoising estimate, additional shots can be assigned to reduce measurement variance. The number of shots should therefore be scheduled according to the step-dependent noise level and the covariance structure of the quantum measurement estimator, rather than fixed conservatively throughout the entire sampling trajectory.

\vspace{-0.25cm}
\subsection{Related Works}
This work lies at the intersection of diffusion-based generative modeling and QC. In the following, we review the related works from both quantum computing and diffusion aspects, and introduce other existing combinations.

QC uses qubits as its basic carriers of information and processes that information in an exponentially large state space through superposition and entanglement. One family of models that is convenient to optimize is the PQC, built from stacked layers of tunable quantum gates and trained by iterating over its parameters much as a neural network is~\cite{benedetti2019pqc,cerezo2021vqa,peruzzo2014}. Unlike a classical network, the output of a quantum circuit cannot be read out directly. Instead the circuit is measured repeatedly in the computational basis, and the probability of each basis state is estimated from a finite number of measurement outcomes, following the Born rule~\cite{nielsen2010}. The readout of a quantum circuit therefore carries an intrinsic sampling noise from this finite measurement, whose magnitude decreases as the number of measurements grows. This sampling noise is the common starting point of later work on measurement cost, and it is the object that we revisit in this paper.

Building on QC, quantum machine learning studies how PQCs can carry out learning tasks~\cite{biamonte2017,schuld2019feature,havlicek2019,mitarai2018}. A central step is how classical data is loaded into a quantum state, for which common encodings include amplitude encoding, angle encoding, and data re-uploading~\cite{perezsalinas2020,larose2020}. The choice of encoding determines not only the form in which data enters the circuit but also the class of functions the circuit can express and how easily it can be trained~\cite{schuld2021encoding,mcclean2018barren,abbas2021power}. A recurring practical obstacle in this area is the readout issue noted above. The circuit output is available only through repeated measurements, the readout fluctuates when the number of measurements is small, and the usual remedy is to take more measurements, which directly raises the computational cost. Our work departs from this common practice. Rather than treating the fluctuation as an error to be suppressed, we attend to its structure and put it to use.

\gthchanged{Parallel to this quantum-computing thread are DMs, including denoising diffusion probabilistic models (DDPM) and score-based generative models~\cite{ddpm,song2019score,song2021sde,nichol2021improved}. Stochastic samplers inject Gaussian noise during reverse sampling. The denoising diffusion implicit model (DDIM) also permits deterministic generation from different random initial states \cite{ddim}. Our method uses the noise prescribed by a stochastic sampler as a budget for finite-shot measurement error.}

Using a PQC as the denoising generator of a DM is a more recent line at this intersection. Quantum generative models have been explored fairly widely, with quantum circuit Born machines and quantum generative adversarial networks as basic architectures~\cite{benedetti2019born,liu2018born,lloyd2018qgan,dallairedemers2018qgan,zoufal2019qgan}, and on this basis quantum and hybrid DMs have appeared more recently~\cite{zhang2024quddpm,parigi2024,kolle2024,cacioppo2023,kwun2024}. Existing quantum diffusion studies mostly realize the forward and reverse processes with a parameterized circuit and assume that high-precision circuit outputs are available, so that the overall measurement cost rises together with the required readout accuracy. Broadly, prior work tends to regard the finite-shot measurement error as an approximation error and suppresses it by taking more measurements. This is the point that our work seeks to handle differently.

Reducing the measurement and sampling cost of QC is therefore a line of research in its own right. Existing schemes include allocating measurements across different observables in variational algorithms, and constructing classical estimates of a quantum state from a small number of measurements~\cite{wecker2015,verteletskyi2020,crawford2021,kubler2020,arrasmith2020,gonthier2022,huang2020shadows}. What these methods share is the goal of keeping the estimation error small with limited measures.

Meanwhile, our starting point is different from both threads above. Generally speaking, the quantum diffusion works assume the requirement of an accurate readout and pay for it. The measurement-frugal works minimize the estimation error under a shot budget. Both therefore treat the error as something to be made small. Instead of minimizing the readout error, we only seek to limit it such that the generation procedure can work normally, making a covariance matching problem instead of minimizing. \gthchanged{Where such condition holds, the sampler reduces the injected-noise covariance by the covariance already contributed by measurement.} The shot count is then an output of the condition rather than a hyperparameter to be tuned, and it is what our adaptive schedule implements.

\vspace{-0.25cm}
\subsection{Contributions}
\vspace{-0.05cm}
The contribution of this work mainly lies on a QC mechanism for diffusion model applications. We turn the finite-shot noise of a quantum readout from a liability into a resource the sampler can spend, and we give the condition that says how far it can be spent. The specific contributions are as follows.
\gthchangebegin
\begin{itemize}
\item We identify the injected-noise covariance of a stochastic diffusion sampler as a budget for finite-shot readout noise, allowing quantum measurement fluctuations to replace part of the externally generated noise.
\item We derive a positive-semidefinite (PSD) condition for matching the total conditional noise covariance to that of the shot-noise-free reverse step, and solve it for a state-dependent shot budget. With the true covariance and conditionally independent residual noise, this gives exact covariance matching for the linear update, which additionally relies on the Gaussian approximation to the readout error.
\item We supply the encoding and the trainable readout that the mechanism needs, placing the diffusion process on a bounded zero-sum hyperplane mapped to the probability simplex so that the Born-measurement output is compatible with the denoising task.
\item We validate the method on a Gaussian mixture and on MNIST in a learned latent space, achieving comparable quality on the reported metrics with far fewer measurements.
\end{itemize}
\gthchangeend

\section{Preliminary}
\subsection{Vanilla Diffusion Model}
The DDPM framework \cite{ddpm} defines a fixed forward noising process that gradually corrupts data and a learned reverse denoising process that synthesizes samples. The distribution flow of the diffusion process can be described as 
\begin{equation}
    \bm{X}_t=\alpha_t\bm{X}_0+\beta_t\bm{W}, \bm{X}_0\sim\mu_0, \bm{W}\sim\mathcal{N}(\bm{0},\mathbf{I}),
\end{equation}
where $\mu_0$ is the target data distribution, $\bm{W}$ is standard Gaussian noise, and $t\in[0,T]$. We adopt a variance-preserving schedule with $\beta_t$ increasing linearly over the $T$ steps,
\begin{equation}
\beta_t=\beta_{\min}+(\beta_{\max}-\beta_{\min})\frac{t}{T},\qquad \alpha_t=\sqrt{1-\beta_t^2}.\label{eq:schedule}
\end{equation}

The corresponding reverse chain is parameterized as a Gaussian distribution,
\begin{equation}
p(\bm{x}_{t^\textrm{next}}|\bm{x}_t)=\mathcal{N}\big(\alpha_{t^\textrm{next}}\mathbb{E}(\bm{x}_0|\bm{x}_t),\beta_{t^\textrm{next}}^2\mathbf{I}\big).
\label{eq:reverse}
\end{equation}
This formulation yields an efficient ancestral sampler by iterating \eqref{eq:reverse} from $t=T$ to $0$ starting at $\mathbf{x}_T\sim\mathcal{N}(\mathbf{0},\mathbf{I})$.

In the practical NN training, we usually only consider discrete time steps, and set the target NN output as the simple noise-prediction objective $\mathbb{E}(\bm{W}|\bm{x}_t)$. Thus, the loss function becomes
\begin{equation}
\mathcal{L}=\mathbb{E}_{t,\bm{x}_0,\bm{\epsilon}}\big\|\bm{\epsilon}-\operatorname{NN}\big(\alpha_t\bm{x}_0+\beta_t\bm{\epsilon},t;\bm{\theta}\big)\big\|_2^2,
\label{eq:loss}
\end{equation}
where $\bm{\theta}$ denotes the NN's parameters.

The reverse process above is the stochastic DDPM sampler. DDIM~\cite{ddim} generalizes it to a family indexed by a parameter $\eta\in[0,1]$ that controls the magnitude of the noise injected at each reverse step,
\begin{equation}
\sigma_t\triangleq\eta\,\frac{\beta_{t^\textrm{next}}}{\beta_t}\sqrt{1-\frac{\alpha^2_t}{\alpha^2_{t^\textrm{next}}}},\qquad\eta\in[0,1],\label{sigma_eta}
\end{equation}
so that $\eta=1$ is the stochastic DDPM sampler above and $\eta=0$ a deterministic one.

\begin{figure}[t]
\centering
\includegraphics[width=0.85\columnwidth]{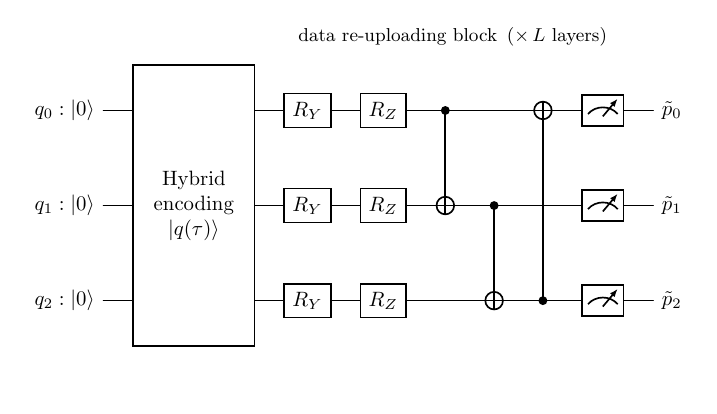}
\vspace{-0.6cm}
\caption{A typical QC based on data re-uploading circuit.}
\label{fig:circuit}
\vspace{-0.2cm}
\end{figure}
\subsection{Basics on Quantum Computing}\label{sec:qc}
In QC, we typically use amplitude encoding or angular encoding, which is realized by rotation gates, to encode the raw input into quantum states. Specifically, if we have an $N$-dimensional input $\bm{x}=(x_1, \cdots, x_N)^\top\in\mathcal{R}^N$, the amplitude encoding becomes
\begin{equation}
\ket{x} = \frac{1}{\|\bm{x}\|_2}\sum_{i=1}^{N}x_i\ket{i},
\end{equation}
where $\ket{i}$ is represented by $\lceil \log_2 N\rceil$ qubits corresponding to the binary representation of $i$.
Meanwhile, the angular encoding is defined as
\begin{equation}
\ket{x} = \bigotimes_{i=1}^N\left(\cos(x_i/2)\ket{0}+\sin(x_i/2)\ket{1}\right),
\end{equation}
where $\bigotimes$ represents the concatenation of quantum states. Thus, angular encoding requires $N$ qubits.
We also have another simple encoding scheme called probability encoding, which is similar to the amplitude encoding that
\begin{equation}
\ket{p}=\sum_{i=1}^{N}\sqrt{p_i}\ket{i},
\end{equation}
which requires that the input $\bm{p}$ is in the probabilistic space $\mathcal{P}^{N-1}\triangleq\{\bm{x}|\bm{x}\in[0,1]^N, \|\bm{x}\|_1=1\}$.
\vspace{-0.45cm}
\subsection{Quantum Neural Networks}
\vspace{-0.05cm}
A QNN is a quantum circuit $U(\bm{\theta})$ whose gate operations depend on a set of trainable parameters $\bm{\theta}$. Its operation can be described in three steps, by analogy with a classical neural network. The input is first encoded into a quantum state $\ket{\psi_{\mathrm{in}}}$. The parameterized circuit then acts on it to give the output state $\ket{\psi_{\mathrm{out}}}=U(\bm{\theta})\ket{\psi_{\mathrm{in}}}$, and a measurement in the computational basis reads out the Born probabilities $p_i=|\braket{i|\psi_{\mathrm{out}}}|^2$ of the basis states. As in a classical network, the parameters $\bm{\theta}$ are trained by gradient descent to minimize a task loss $\mathcal{L}(\bm{\theta})$, which for the diffusion denoiser of this paper is specified in Section~\ref{sec:train}.

A common circuit structure stacks $L$ layers of identical form. As shown in Fig.~\ref{fig:circuit}, each layer consists of two parts. The first is a set of single-qubit parameterized rotations $R_Y(\theta)=e^{-i\theta Y/2}$ and $R_Z(\theta)=e^{-i\theta Z/2}$, with $Y$ and $Z$ the Pauli matrices, applied to every qubit. The second is a fixed entangling block, for instance a ring of CNOT gates that connects the qubits end to end, which correlates the qubits. Writing $\bm{V}_l(\bm{\theta}_l)$ for the combined rotation-and-entangling operation of layer $l$, the circuit is their composition $U(\bm{\theta})=\bm{V}_L(\bm{\theta}_L)\cdots\bm{V}_1(\bm{\theta}_1)$. To give a circuit of modest depth enough expressive power, the input features can be re-injected into the rotation angles of every layer, a technique known as data re-uploading~\cite{perezsalinas2020,schuld2021encoding}. By feeding the same input into different layers repeatedly, a shallow circuit can represent more complex functions.

In this work, we consider the implementation of DM with a QNN, whose structure is illustrated in Fig.~\ref{fig:circuit}.
The QNN takes the encoded noisy sample and the current diffusion step as input. The input is first loaded into $\ket{q(\tau)}$ with time embedding through quantum encoding, whose details are discussed in the following section. After $L$ layers of rotations and CNOT rings as in Fig.~\ref{fig:circuit}, the result is measured in the computational basis to give the estimate $\tilde{\bm{p}}$ of the probability. A readout layer then decodes $\tilde{\bm{p}}$ into a prediction of the clean signal. It is a classical linear layer applied to the measurement outcomes, which is not a part of the quantum circuit.

\vspace{-0.25cm}
\section{Diffusion Model on Probabilistic Space}
\begin{figure*}[t]
\centering
\includegraphics[width=0.9\textwidth]{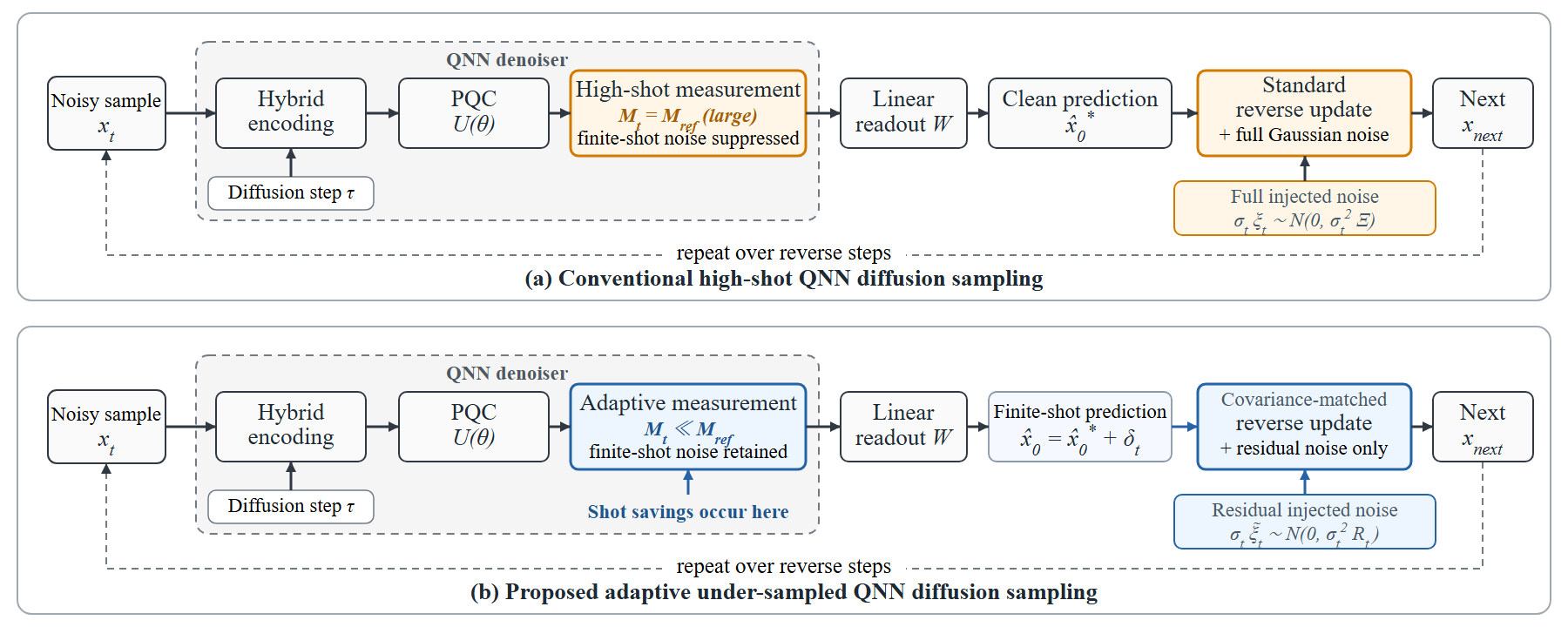}
\caption{\gthchanged{QNN diffusion sampling with (a) conventional high-shot readout and (b) adaptive under-sampling that reuses measurement noise.}}
\label{fig:pipeline}
\vspace{-0.25cm}
\end{figure*}

\gthchanged{Figure~\ref{fig:pipeline} compares two sampling pipelines for quantum diffusion models using the same QNN denoiser and classical readout. At each reverse step, the QNN predicts the clean sample from the current noisy state and diffusion time. Conventional sampling uses many shots to suppress measurement noise, then adds the full prescribed diffusion noise. Our sampler uses fewer shots and reduces the added noise to account for the measurement fluctuations. Both start from random noise: denoising is the prediction at each step, and generation is the complete reverse loop.}

\vspace{-0.3cm}
\subsection{Hybrid Encoding for Quantum Diffusion Models}\label{sec:enc}
For a DM, the encoding is not a free design choice, and we begin by making the constraint that fixes it explicit. A DM applies its denoiser iteratively. The output of one reverse step is the input of the next, so the encoding and the readout must live in the same space. On the quantum side the readout is fixed by physics, since measuring in the computational basis returns Born probabilities and the output of the circuit therefore always lies in the simplex $\mathcal{P}^{N-1}$. The input must consequently be placed on the simplex as well, and probability encoding, which loads $\sqrt{p_i}$ into the amplitude of $\ket{i}$, is the encoding that the problem selects.

The other two encodings of Section~\ref{sec:qc} play different roles here. Angle encoding uses one qubit per data dimension, and we adopt it in Section~\ref{sec:8q}, where a low data dimension makes it affordable. Amplitude encoding is more expressive per qubit, but its output is a probability vector while its input is a signed amplitude vector, so the two do not live in the same space and the denoiser cannot be iterated without an extra and costly reconstruction step.

Probability encoding leaves one resource untouched. It fixes only the moduli of the amplitudes and says nothing about their phases, so the azimuthal degrees of freedom of the state are unused. A DM, in turn, must supply its denoiser with one quantity beyond the noisy sample, namely the current diffusion step, which is conventionally injected through a time embedding. The information the denoiser has to carry is thus exactly the information the encoding has room for. We therefore write the diffusion step into the phases, which gives the following hybrid encoding of probability and time,
\begin{equation}
\ket{p(t)}\triangleq\sum_{i=1}^{N}\sqrt{p_i}e^{jf_i(t)}\ket{i},
\end{equation}
where $\bm{p}\in\mathcal{P}^{N-1}$ is the input and $\bm{f}(t)$ is the time embedding. The probability encoding requires the entries of $\bm{p}$ to sum to one, so an $n_q$-qubit state, which carries $N=2^{n_q}$ probabilities, has only $N-1=2^{n_q}-1$ independent degrees of freedom. The generative task is therefore represented on an $(N-1)$-dimensional space.

We write the diffusion step into the phases by letting the phase of basis state $i$ advance linearly in the normalized step $\tau\triangleq t/(T-1)\in[0,1]$,
\begin{equation}
f_i(\tau)\triangleq c\,w_i\,\tau,\qquad w_i\triangleq a\,i+b\,i^{2},\qquad i=1,\dots,N,\label{timeemb}
\end{equation}
with $a,b>0$ shaping the rates across the basis and $c>0$ setting their overall scale. We fix that scale so the largest phase excursion $\phi_{\max}$ stays within one revolution,
\begin{equation}
c=\frac{\phi_{\max}}{aN+bN^{2}},\qquad \phi_{\max}\in(0,2\pi],\label{cnorm}
\end{equation}
so that the embedding is controlled by the single scalar $\phi_{\max}$, while $a$ and $b$ only redistribute the phase budget across the basis.

\begin{figure*}[t]
\centering
\vspace{-0.4cm}
\includegraphics[width=0.9\textwidth]{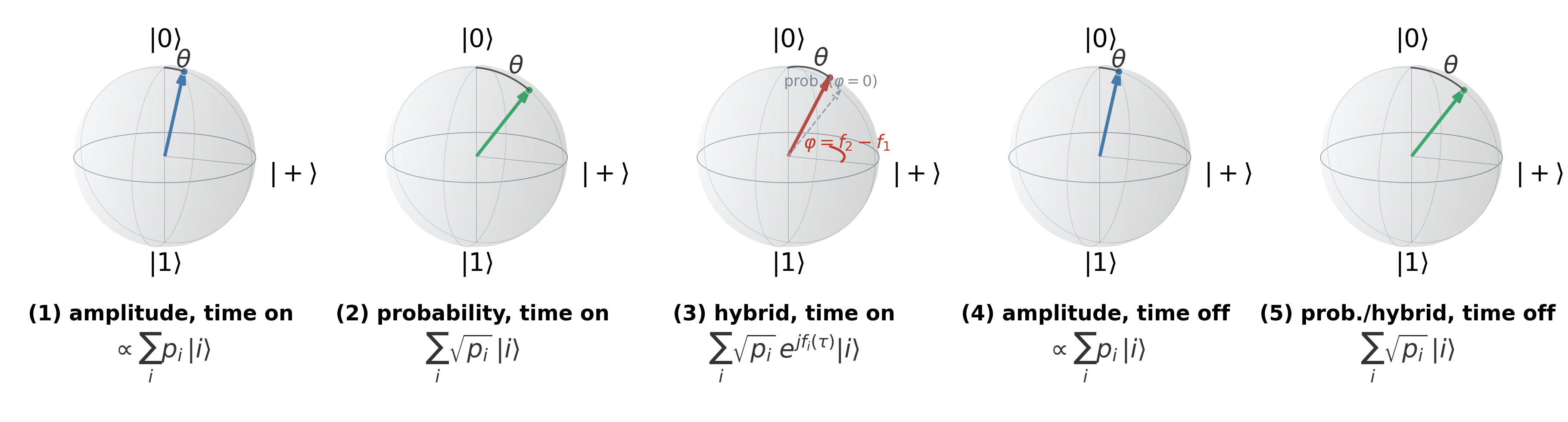}
\vspace{-0.6cm}
\caption{Encoding schemes on the Bloch sphere, shown with a single-qubit illustration.}
\vspace{-0.2cm}
\label{fig:encoding}
\end{figure*}

Figure~\ref{fig:encoding} illustrates these encodings on the Bloch sphere, using a single-qubit input for clarity while the actual system uses $n_q$ qubits. Amplitude encoding and probability encoding place the state at different polar angles, using the probability values and their square roots as amplitudes respectively, and a natural property of the latter is that the Born probabilities obtained from measurement equal the input distribution. The hybrid encoding builds on probability encoding and writes the diffusion step into the azimuthal phase. Switching off the time embedding removes that phase, so the fourth and fifth schemes in Fig.~\ref{fig:encoding} coincide with the first and the second.

\vspace{-0.35cm}
\subsection{Diffusion Model on the Zero-Sum Hyperplane}\label{sec:diff}
The encoding above operates on the probability simplex, which is not where practical data lives. To bridge the two, we place the data on the bounded zero-sum hyperplane
\begin{equation}
    \bm{x}\in\mathcal{S}^{N-1}(A)\triangleq\{\bm{x}|\bm{x}\in[-A,A]^N, \bm{1}^\top\bm{x}=0\},\label{restriction}
\end{equation}
and denote the orthogonal projection matrix onto its unbounded version $\mathcal{S}^{N-1}(\infty)$ by
\begin{equation}
\mathbf{\Xi}\triangleq\mathbf{I}-\frac{1}{N}\bm{1}\bm{1}^\top,
\end{equation}
so that the standard Gaussian on the hyperplane becomes $\mathcal{N}(\bm{0},\mathbf{\Xi})$. With appropriate preprocessing a reasonable bound $A$ can be obtained and the mean of each sample can be set to zero. Where the bias matters, one further dimension can be added to $\bm{x}$ to absorb it.

It is worth being explicit about how a task is placed on the hyperplane, because the two natural ways of doing so are not equivalent. The constraint $\bm{1}^\top\bm{x}=0$ removes one degree of freedom, so the hyperplane is an $(N-1)$-dimensional subspace, and the circuit sees $N$ coordinates while the task carries only $K\triangleq N-1$ of them. Let the columns of $\mathbf{U}\in\mathbb{R}^{N\times K}$ be an orthonormal basis of the hyperplane, that is, the eigenvectors of $\mathbf{\Xi}$ with eigenvalue one, so that
\begin{equation}
\mathbf{U}^\top\mathbf{U}=\mathbf{I}_K,\qquad \mathbf{U}\mathbf{U}^\top=\mathbf{\Xi},\qquad \bm{1}^\top\mathbf{U}=\bm{0}^\top.\label{basis}
\end{equation}
A task described by $K$ free coordinates $\bm{u}\in\mathbb{R}^{K}$ is then lifted onto the hyperplane by
\begin{equation}
\bm{x}=\mathbf{U}\bm{u},\qquad\text{with exact inverse}\qquad \bm{u}=\mathbf{U}^\top\bm{x},\label{lift}
\end{equation}
a bijection between $\mathbb{R}^{K}$ and the hyperplane. The zero-sum constraint holds by construction and costs no information, since the $N$ coordinates handed to the circuit carry precisely the $K$ degrees of freedom of the task. The alternative, taking $N$ coordinates and projecting them by $\bm{x}\mapsto\mathbf{\Xi}\bm{x}$, also lands on the hyperplane but is not injective. It discards the component along $\bm{1}$, which is then not recoverable at decoding time. The distinction is immaterial when the data is natively zero-sum, but it matters as soon as the $N$ coordinates are produced by a dimensionality-reducing code, in which case the code should be asked for $K=N-1$ coordinates and lifted by \eqref{lift} rather than for $N$ coordinates and projected. We use the lifted form throughout.

Under \eqref{restriction}, a hyperplane point $\bm{\chi}\in\mathcal{S}^{N-1}(A)$ and a probability vector $\bm{p}\in\mathcal{P}^{N-1}$ are related by
\begin{align}
    \bm{p} &= \frac{1}{NA}\bm{\chi}+\frac{1}{N}&(\mathcal{S}^{N-1}(A)\rightarrow\mathcal{P}^{N-1}),\label{forwardmap}\\
    \bm{\chi} &= A\operatorname{clamp}_{[-1,1]}(N\bm{p}-1)&(\mathcal{P}^{N-1}\rightarrow\mathcal{S}^{N-1}(A)),\label{backmap}
\end{align}
where $\operatorname{clamp}_{[-1,1]}(\cdot)$ clamps each element to $[-1,1]$. The forward map is exactly the affine bijection announced above, and it is the reason the Born probabilities of the circuit can be compared with the data directly.

Following the vanilla procedure on $\mathcal{S}^{N-1}(\infty)$, the forward process of the DM reads
\begin{equation}
    \bm{X}_t=\alpha_t\bm{X}_0+\beta_t\bm{W}, \bm{X}_0\sim\mu_0, \bm{W}\sim\mathcal{N}(\bm{0},\mathbf{\Xi}).\label{diff}
\end{equation}
We adopt the $\bm{x}_0$-prediction parameterization, in which the denoiser directly estimates the clean signal, with training target
\begin{equation}
\bm{g}(\bm{x}_t, t)\triangleq\mathbb{E}(\bm{X}_0|\bm{X}_t=\bm{x}_t), \label{target}
\end{equation}
the expectation being taken under the prior in \eqref{diff}. Predicting the noise $\bm{W}$ instead would reconstruct the clean signal as $(\bm{x}_t-\beta_t\bm{g})/\alpha_t$ and amplify the readout error by $1/\alpha_t$ in the early, high-noise steps. Predicting $\bm{X}_0$ avoids that amplification, which matters here because the readout is bounded, so that an error divided by a small $\alpha_t$ would be driven outside the range the readout can represent.

The reverse step is the DDIM update \cite{ddim} written on the hyperplane. Writing $\hat{\bm{x}}_0$ for the estimate of the clean signal, the noise direction implied by the current iterate is $\hat{\bm{\varepsilon}}_t=(\bm{x}_t-\alpha_t\hat{\bm{x}}_0)/\beta_t$, and
\begin{equation}
\bm{x}_{t^\textrm{next}}=\alpha_{t^\textrm{next}}\hat{\bm{x}}_0+\sqrt{\beta^2_{t^\textrm{next}}-\sigma_t^2}\;\hat{\bm{\varepsilon}}_t+\sigma_t\bm{\xi}, \label{denoising}
\end{equation}
where $\bm{\xi}\sim\mathcal{N}(\bm{0},\mathbf{\Xi})$ and $\sigma_t$ is the injected-noise scale of \eqref{sigma_eta}. It is convenient to collect the coefficient multiplying the denoiser output,
\begin{equation}
k_t\triangleq\alpha_{t^\textrm{next}}-\frac{\alpha_t}{\beta_t}\sqrt{\beta^2_{t^\textrm{next}}-\sigma_t^2},\label{kt}
\end{equation}
\gthchanged{so that \eqref{denoising} becomes $\bm{x}_{t^\textrm{next}}=k_t\hat{\bm{x}}_0+\big(\sqrt{\beta^2_{t^\textrm{next}}-\sigma_t^2}/\beta_t\big)\bm{x}_t+\sigma_t\bm{\xi}$. Two quantities drive the whole analysis of Section~\ref{sec:algo}. They are $\sigma_t$, the noise scale prescribed by the chosen stochastic reverse sampler, and $k_t$, the factor by which an error in the clean-signal estimate, including finite-shot readout error, propagates into the next iterate. The term $\sigma_t\bm{\xi}$ is added by hand, and it is precisely this deliberate randomness that we reuse to absorb the noise of an under-sampled readout, instead of paying measurements to suppress it.}

\makeatletter
\renewcommand{\section}{\@startsection{section}{1}{\z@}{2ex plus 1ex minus .5ex}{.7ex plus .5ex}{\normalfont\normalsize\centering\scshape}}
\renewcommand{\subsection}{\@startsection{subsection}{2}{\z@}{2ex plus 1ex minus .5ex}{.7ex plus .5ex}{\normalfont\normalsize\itshape}}
\makeatother
\subsection{Readout}\label{sec:readout}
The encoding, the circuit, and the fixed decoding \eqref{backmap} already form a complete neural network, but its compression ratio is too high, so we raise its capacity with a trainable linear readout, a classical linear layer applied after the measurement,
\begin{equation}
\hat{\bm{x}}_0=\mathbf{\Xi}\big(\mathbf{W}^\top\tilde{\bm{p}}+\bm{b}\big),\label{eq:readout}
\end{equation}
\gthchanged{where $\mathbf{W}\in\mathbb{R}^{N\times N}$ and $\bm{b}\in\mathbb{R}^{N}$ are trained jointly with the circuit parameters, and $\mathbf{\Xi}$ keeps the estimate on the hyperplane. Setting $\mathbf{W}=AN\mathbf{I}$ and $\bm{b}=-A\bm{1}$ recovers the affine inverse of \eqref{forwardmap}. It agrees with \eqref{backmap} when the clamp is inactive. We use these affine parameters to initialize the trainable readout. Unless otherwise stated, all results in this paper use the trainable readout \eqref{eq:readout}.}

We note that \eqref{eq:readout} is classical post-processing that acts on the measurement outcomes, outside the quantum circuit, and adds no quantum cost. We also note that, less obviously, it is not neutral with respect to measurement cost. Because it multiplies the finite-shot fluctuation of $\tilde{\bm{p}}$, it reshapes the covariance of that fluctuation, and the number of shots the sampler needs follows from that covariance rather than from the raw multinomial one. Section~\ref{sec:algo} makes this precise, and Section~\ref{sec:gmm_readout} shows that the effect is large.

The forward map \eqref{forwardmap} returns a valid probability vector only when its argument lies in the box $[-A,A]^N$, while the reverse process is not confined to the box a priori. We therefore treat $\mathcal{S}^{N-1}(A)$ as the feasible set of the sampler and project each iterate onto it, so that \eqref{forwardmap} remains well defined at the next step. The readout \eqref{eq:readout} is a purely linear map with no clamp, so the clamp of \eqref{backmap} survives only in the fixed-readout special case.

\subsection{Training}\label{sec:train}
\gthchanged{For the fixed affine readout, the ideal denoising target \eqref{target} corresponds to the probability vector}
\begin{equation}
\bar{\bm p}_t\triangleq \frac{1}{NA}\mathbb{E}(\bm{X}_0|\bm{X}_t=\bm{x}_t)+\frac{1}{N},\label{eq:ptarget}
\end{equation}
\gthchanged{with circuit input $\frac{1}{NA}\bm{x}_t+\frac{1}{N}$ by \eqref{forwardmap}. With a trainable readout, the loss constrains the decoded clean-signal prediction rather than requiring the circuit probabilities to equal this vector. Training then follows the standard denoising recipe. We draw $\bm{x}_0$ from the data and $t$ uniformly, form $\bm{x}_t=\alpha_t\bm{x}_0+\beta_t\bm{\xi}$ with $\bm{\xi}\sim\mathcal{N}(\bm{0},\mathbf{\Xi})$, map it to the simplex by \eqref{forwardmap}, and minimize the squared error between the readout and the target,}
\begin{equation}
\mathcal{L}(\bm{\theta},\mathbf{W},\bm{b})=\big\|\hat{\bm x}_0-\bm{x}_0\big\|_2^2,\label{eq:qloss}
\end{equation}
over the circuit parameters $\bm{\theta}$ and the readout $(\mathbf{W},\bm{b})$ jointly. Note that \eqref{eq:qloss} predicts $\bm{x}_0$ and not the noise, unlike the vanilla objective \eqref{eq:loss}, and that the encoding leaves no choice in the matter. The conditional mean $\mathbb{E}(\bm{X}_0|\bm{X}_t)$ lies in the box $[-A,A]^N$ and is therefore carried onto the simplex by \eqref{forwardmap}, whereas the noise is unbounded and has no image there. Training is carried out at exact probabilities. The finite-shot analysis of the following sections concerns generation only.

\section{Under-Sampling Quantum DDPM Denoising Algorithm}\label{sec:algo}
The output of a quantum circuit is available only through repeated measurement, so its readout carries a sampling error of a definite structure. We first characterize that error, then show how the randomness the reverse step injects anyway can absorb it, and finally determine how many measurements this leaves us free to save.

\gthchangebegin
\begin{theorem}\label{the:proberr}
Consider an $N$-valued discrete random variable $V$ which takes the values $v_i$ with probabilities $p_i$, $i=1,\cdots,N$, respectively. With $M$ repeated observations, we have a group of independent samples of $V$, denoted by $V_1,\cdots,V_M$. Denote $\bm{p}\triangleq(p_1,\cdots,p_N)^\top$ and $\tilde{\bm{p}}\triangleq(\tilde{p}_1,\cdots,\tilde{p}_N)^\top$, where $\tilde{p}_i\triangleq\frac{1}{M}\sum_{j=1}^{M}\mathbbm{1}_{V_j=v_i}$ indicates the estimated probability through observations.

The estimator satisfies $\mathbb{E}(\tilde{\bm p}-\bm p)=\bm0$ and $\operatorname{Cov}(\tilde{\bm p}-\bm p)=\mathbf\Sigma/M$ exactly. For fixed $\bm p$, the normalized central limit theorem (CLT) motivates the finite-shot Gaussian approximation
\begin{equation}
    \tilde{\bm{p}}-\bm{p}\mathrel{\overset{\mathrm{approx.}}{\sim}}\mathcal{N}(\bm{0}, \frac{1}{M}\mathbf{\Sigma}),\label{err}
\end{equation}
where
\begin{equation}
\begin{aligned}
\Sigma_{i,j}&=p_i(1-p_i),& i=j,\\
\Sigma_{i,j}&=-p_ip_j.& i\neq j
\end{aligned}\label{sigma}
\end{equation}
\end{theorem}
\gthchangeend
\gthchangebegin
\begin{proof}
Define random variable $\bm{e}\triangleq(\mathbbm{1}_{V=v_1}, \cdots,\mathbbm{1}_{V=v_N})^\top$. We easily know $\mathbb{E}(\bm{e})=\bm{p}$ and $\operatorname{Cov}(\bm{e})=\mathbf{\Sigma}$, where $\mathbf{\Sigma}$ is defined in \eqref{sigma}.
With a group of independent samplings $\bm{e}_1,\cdots,\bm{e}_M$, we know
\begin{equation}
\sqrt{M}(\frac{1}{M}\sum_{m=1}^{M}\bm{e}_m-\bm{p})\xrightarrow{d}\mathcal{N}(\bm{0}, \mathbf{\Sigma}).
\end{equation}
The mean and covariance follow exactly from independence of the observations. The normalized convergence above motivates the approximation in \eqref{err}, since $\tilde{\bm p}=M^{-1}\sum_{m=1}^{M}\bm e_m$.
\end{proof}
\gthchangeend

\gthchangebegin
\begin{assumption}[Linear-update comparison and ideal denoiser]\label{ass}
Throughout the analysis of the reverse process we assume that
\begin{itemize}
\item[(i)] the moment comparison concerns the linear reverse update before feasibility projection onto $\mathcal{S}^{N-1}(A)$,
\item[(ii)] the denoiser is trained to its optimum, so that the shot-noise-free readout matches the target, $\mathbf{\Xi}(\mathbf{W}^\top\bm{p}+\bm{b})=\bm{g}(\bm{x}_t,t)$.
\end{itemize}
\end{assumption}
\gthchangeend

\gthchanged{The moment calculation applies before projection, which can change the moments of the update. A finite bound $A$ does not exclude Gaussian boundary excursions. Part (ii) identifies the shot-noise-free readout with the ideal conditional mean; comparison with the same trained denoiser does not require optimal training.}

\gthchanged{At each state, the calculation uses a fixed shot count, the true readout covariance, and residual noise conditionally independent of the measurement outcomes. These conditions give exact conditional mean and covariance matching for the linear update. Distribution matching additionally uses the Gaussian approximation.}

\subsection{Absorbing the Finite-Shot Noise}\label{sec:absorb}
\gthchanged{Passing the finite-shot error of Theorem~\ref{the:proberr} through the linear readout \eqref{eq:readout} gives}
\gthchangebegin
\begin{equation}\label{sample}
\hat{\bm{x}}_0=\hat{\bm{x}}_0^{\star}+\bm{\delta},\qquad \bm{\delta}\mathrel{\overset{\mathrm{approx.}}{\sim}}\mathcal{N}\Big(\bm{0},\ \tfrac{1}{M}\,\mathbf{\Xi}\mathbf{W}^\top\mathbf{\Sigma}\mathbf{W}\mathbf{\Xi}\Big),
\end{equation}
\gthchangeend
\gthchanged{Here $\hat{\bm{x}}_0^{\star}=\mathbf{\Xi}(\mathbf{W}^\top\bm{p}+\bm{b})$ is the shot-noise-free readout, equal to $\bm{g}(\bm{x}_t,t)$ under Assumption~\ref{ass}. The mean and covariance are exact; normality is approximate. The readout reshapes the multinomial covariance through $\mathbf W$.}

Substituting \eqref{sample} into the reverse step \eqref{denoising} and using the coefficient $k_t$ of \eqref{kt}, the readout error enters the next iterate as $k_t\bm{\delta}$, alongside the injected noise $\sigma_t\bm{\xi}$. Instead of suppressing $\bm{\delta}$ by taking more measurements, we subtract its contribution from the noise we were going to inject. We replace $\bm{\xi}\sim\mathcal{N}(\bm{0},\mathbf{\Xi})$ by $\tilde{\bm{\xi}}\sim\mathcal{N}(\bm{0},\mathbf{R}_t)$ with
\begin{align}
\bm{x}_{t^\textrm{next}}&=\alpha_{t^\textrm{next}}\hat{\bm{x}}_0+\sqrt{\beta^2_{t^\textrm{next}}-\sigma^2_t}\;\hat{\bm{\varepsilon}}_t+\sigma_t\tilde{\bm{\xi}},\label{adjusted_ddpm}\\
\mathbf{R}_t&\triangleq\mathbf{\Xi}-\frac{k_t^2}{\sigma_t^2 M}\,\mathbf{\Xi}\mathbf{W}^\top\mathbf{\Sigma}_t\mathbf{W}\mathbf{\Xi},\label{adjusted_sample}
\end{align}
\gthchanged{so that $\operatorname{Cov}(k_t\bm{\delta}+\sigma_t\tilde{\bm{\xi}})=\sigma_t^2\mathbf{\Xi}$, matching the prescribed noise covariance. The residual covariance must be positive semidefinite:}
\begin{equation}
\mathbf{\Xi}-\frac{k_t^2}{\sigma_t^2 M}\,\mathbf{\Xi}\mathbf{W}^\top\mathbf{\Sigma}_t\mathbf{W}\mathbf{\Xi}\succeq 0.\label{requirement}
\end{equation}
\gthchanged{Condition~\eqref{requirement} is the central object of this paper. It states that the covariance of the sampling noise, after being shaped by the readout and amplified by $k_t$, must not exceed the covariance of the noise budgeted by the reverse step. Both matrices are supported on the zero-sum hyperplane, since $\mathbf{\Xi}$ has eigenvalue $0$ along $\bm{1}$ and $1$ on its orthogonal complement, and $\mathbf{\Sigma}$ also annihilates $\bm{1}$. Hence \eqref{requirement} is a condition on the same $(N-1)$-dimensional subspace and can be satisfied by sufficiently many shots when $\sigma_t>0$, provided that the required count is not truncated by the upper cap.}

\subsection{The Adaptive Shot Budget}\label{sec:budget}
Because \eqref{requirement} is a Loewner-order condition on that subspace, it is equivalent to a scalar bound on $M$ through the largest eigenvalue of the shaped covariance. This yields the per-step measurement budget
\begin{equation}
M_t=\operatorname{clamp}_{[M_\mathrm{min}, M_\mathrm{max}]}\left(\left\lceil\frac{k_t^2}{\sigma_t^2}\,\lambda_{\max}\!\big(\mathbf{\Xi}\mathbf{W}^\top\mathbf{\Sigma}_t\mathbf{W}\mathbf{\Xi}\big)\right\rceil\right),\label{M}
\end{equation}
\gthchanged{where $\mathbf{\Sigma}_t$ is evaluated at the probabilities of the current step and $[M_\mathrm{min},M_\mathrm{max}]$ sets a lower measurement safeguard and a practical resource cap. Equation~\eqref{M} is what our sampler implements. The budget is recomputed at every reverse step, and it depends on the current state through $\mathbf{\Sigma}_t$, on the sampler through $k_t/\sigma_t$, and on the readout through $\mathbf{W}$.}

Two consequences of \eqref{M} are worth stating explicitly.

\gthchanged{Equation~\eqref{M} is evaluated at the current state, which raises a question about what a device can actually do. Reading $\bm{p}$ requires measurement, and how much to measure is the quantity being decided, so the budget as written is circular on hardware even though a simulator resolves it for free. Three ways out are available and none of them changes the mechanism. The first is a pilot batch of $M_\mathrm{min}$ shots whose estimate evaluates $\mathbf{\Sigma}_t$ before the remaining shots are allocated, with the pilot counted in the total. This is cheap here because the budget already sits at $M_\mathrm{min}$ over the early steps, where the pilot is the whole allocation. The second is to reuse the estimate of the preceding reverse step, which costs nothing extra and is accurate because the trajectory varies slowly. The third removes the state dependence altogether. For an arbitrary probability vector, $\|\mathbf\Sigma_t\|_2\le1/2$: the absolute row sum of $\operatorname{Diag}(\bm p)-\bm p\bm p^\top$ is $2p_i(1-p_i)\le1/2$. Consequently $\lambda_{\max}(\mathbf\Xi\mathbf W^\top\mathbf\Sigma_t\mathbf W\mathbf\Xi)\le\|\mathbf W\|_2^2/2$, giving a conservative state-independent budget. The tighter probability bound in Corollary~\ref{cor:fixed} uses the fixed affine readout and must not be assumed for an unrestricted trainable readout. The results reported below use \eqref{M} evaluated at the simulator probabilities, which isolates the absorption mechanism from the cost of estimating $\mathbf{\Sigma}_t$.}

\gthchangebegin
\begin{corollary}[Fixed readout]\label{cor:fixed}
Freezing the readout at $\mathbf{W}=AN\mathbf{I}$, $\bm{b}=-A\bm{1}$ recovers the unclipped affine inverse of \eqref{forwardmap} and reduces \eqref{M} to a closed form. Indeed $\mathbf{\Xi}\mathbf{W}^\top\mathbf{\Sigma}\mathbf{W}\mathbf{\Xi}=(AN)^2\,\mathbf{\Xi}\mathbf{\Sigma}\mathbf{\Xi}$, and by the secular equation for rank-one perturbations of diagonal matrices \cite{bunch1978rank} the nonzero eigenvalues of $\mathbf{\Sigma}=\operatorname{Diag}(\bm{p})-\bm{p}\bm{p}^\top$ interlace with the entries $\{p_i\}$, so that $\|\mathbf{\Sigma}\|_2\le\|\bm{p}\|_\infty$. If this fixed readout equals the bounded target in Assumption~\ref{ass}(ii), its inverse relation gives $p_i=(g_i/A+1)/N$ and hence $\|\bm p\|_\infty\le2/N$. Hence
\begin{equation}
M_t\le\operatorname{clamp}_{[M_\mathrm{min}, M_\mathrm{max}]}\left(\left\lceil\frac{2N A^2 k_t^2}{\sigma_t^2}\right\rceil\right),\label{Mfixed}
\end{equation}
a bound that no longer depends on the current state, and which recovers the closed-form budget of a fixed, parameter-free decoding.
\end{corollary}
\gthchangeend

\gthchangebegin
\begin{corollary}[Cost of determinism]\label{cor:eta}
By \eqref{sigma_eta}, $\sigma_t$ is proportional to $\eta$, while $k_t\to\alpha_{t^\textrm{next}}-\alpha_t\beta_{t^\textrm{next}}/\beta_t$ as $\eta$ varies over $[0,1]$ only through the second-order term $\sigma_t^2$ in \eqref{kt}. Consequently
\begin{equation}
M_t\ \propto\ \frac{1}{\eta^{2}}\label{eta_cost}
\end{equation}
to leading order. As the injected-noise budget tends to zero, the shot count required by this covariance-compensation rule diverges for nonzero readout variance. This is a limitation of absorbing measurement noise into a prescribed stochastic update, not a requirement that deterministic diffusion sampling use infinitely many shots to produce useful samples. The experimentally observed cost trend favors the fully stochastic setting under the schedules studied here. This is the quantitative form of the mechanism this paper proposes. The randomness that the reverse process injects is a budget, and a sampler that injects more of it can be measured less.
\end{corollary}
\gthchangeend

\begin{figure*}[t]
\centering
\includegraphics[width=0.9\textwidth]{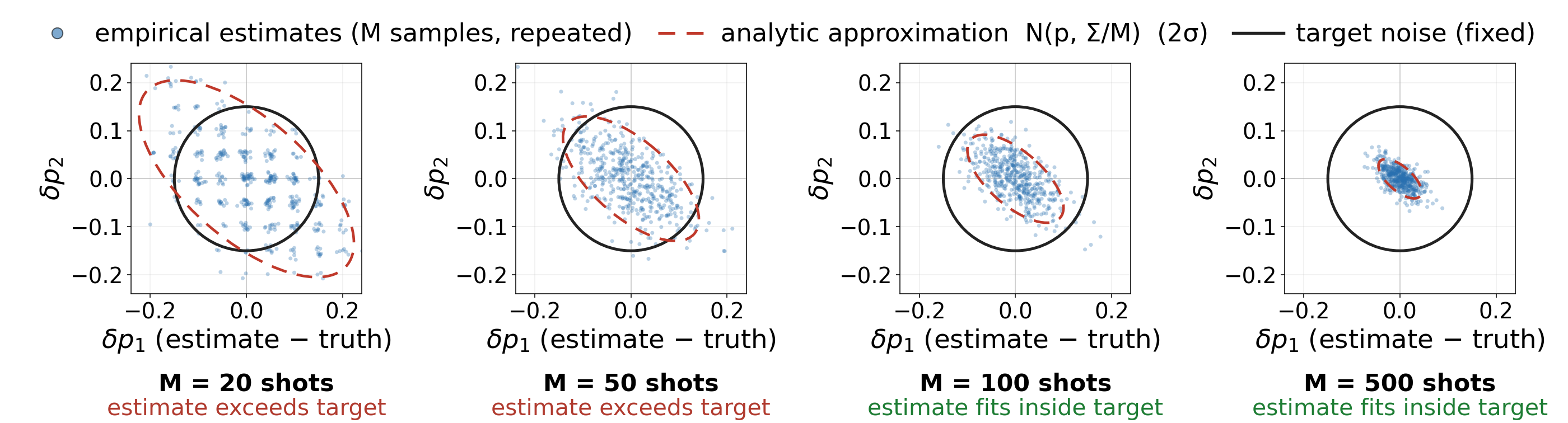}
\caption{Finite-shot sampling approximation on a low-dimensional example.}
\label{fig:sampling}
\end{figure*}

Figure~\ref{fig:sampling} illustrates the absorption on a three-outcome example, plotting the deviation of the estimate from the true value. The cloud of points is the actual sampling noise at a given number of measurements $M$, and the red dashed ellipse is the asymptotic Gaussian approximation $\bm{\Sigma}/M$ of Theorem~\ref{the:proberr}. Across the panels $M$ increases, so the ellipse deviates from the cloud when $M$ is small and matches it when $M$ is large. The black circle is the noise the reverse step injects, whose size is set by the schedule and is independent of $M$. When the cloud lies inside the circle the measurement noise is absorbed at no extra cost, whereas an $M$ too small lets the cloud spill out and biases the generated distribution. The figure is drawn in the probability space for clarity, while the exact comparison~\eqref{requirement} is carried out in the data space, where the two-dimensional picture carries over because the multinomial covariance is rank-deficient in the same way in any dimension.

\begin{remark}[Shape of the schedule]
The budget \eqref{M} is governed by the ratio $k_t/\sigma_t$. Early in the reverse process the injected noise $\sigma_t$ is large and $M_t$ falls back on the lower bound $M_\mathrm{min}$. As the trajectory proceeds, $\sigma_t$ shrinks and $M_t$ grows, so the measurements are concentrated in the final steps where the sampler is close to the data. The schedule is thus increasing along the trajectory, and the saving with respect to a uniform allocation comes precisely from the early steps, in which a high-accuracy readout would be wasted.
\end{remark}

\gthchanged{Algorithm~\ref{alg:ddpm} samples the residual noise using the PSD square root of $\mathbf R_t$. Negative eigenvalues are clipped to zero. If the shot cap leaves the measurement variance above the target in a direction, clipping removes the external noise there but cannot remove the measurement error. Section~\ref{sec:gmm_bounds} examines this effect.}

\gthchanged{The algorithm estimates the residual covariance from the same measured probabilities used for readout. This plug-in estimate introduces sampling dependence beyond the ideal calculation above; its effect is evaluated through the generation experiments.}
\begin{algorithm}[t]
\gthchangebegin
\caption{Under-Sampled Quantum DDPM Reverse Process}
\label{alg:ddpm}
\setlength{\linewidth}{\dimexpr\linewidth-\algomargin\relax}
\begin{algorithmic}[1]
\REQUIRE Trained QNN $\operatorname{NN}(\cdot,\cdot;\bm{\theta})$, readout $(\mathbf{W},\bm{b})$, reverse-step stride $s$, schedule $\{\alpha_t\},\{\beta_t\}$, stochasticity $\eta$, bounds $[M_\mathrm{min},M_\mathrm{max}]$
\STATE Sample $\bm{\xi}\sim\mathcal{N}(\bm{0},\mathbf{I})$, $\bm{x}\leftarrow\mathbf{\Xi}\bm{\xi}$, $t\leftarrow T$\quad\,\COMMENT{Initiation}
\WHILE{$t>0$}
\STATE $t^\textrm{next}\leftarrow \max\{t-s,0\}$
\STATE Compute $\sigma_t$ by \eqref{sigma_eta} and $k_t$ by \eqref{kt}
\STATE Map $\bm{x}$ to the simplex: $\bm{q}\leftarrow\frac{1}{NA}\bm{x}+\frac{1}{N}$
\STATE Obtain $\mathbf{\Sigma}_t$ and get $M_t$ by \eqref{M}\quad\COMMENT{See Section~\ref{sec:budget}}
\STATE Measure $\tilde{\bm p}=\operatorname{NN}(\bm{q},t;\bm{\theta})$ with $M_t$ shots
\STATE Read out $\hat{\bm x}_0\leftarrow\mathbf{\Xi}(\mathbf{W}^\top\tilde{\bm p}+\bm b)$ by \eqref{eq:readout}
\STATE $\mathbf{\Sigma}_t\leftarrow\operatorname{diag}(\tilde{\bm p})-\tilde{\bm p}\tilde{\bm p}^\top$\quad\COMMENT{Residual noise}
\STATE Eigendecompose $\mathbf{R}_t$ of \eqref{adjusted_sample} as $\mathbf{U\Lambda U}^\top$
\STATE $\Lambda_{i,i}\leftarrow\max\{\Lambda_{i,i},0\}$\quad\COMMENT{Enforce PSD for the estimated covariance}
\STATE Sample $\bm{\zeta}\sim\mathcal{N}(\bm{0},\mathbf{I})$, set $\tilde{\bm\xi}\leftarrow\mathbf{U}\mathbf{\Lambda}^{1/2}\bm{\zeta}$\quad\COMMENT{$\tilde{\bm\xi}\sim\mathcal{N}(\bm{0},\mathbf{R}_t)$}
\STATE $\hat{\bm\varepsilon}_t\leftarrow(\bm{x}-\alpha_t\hat{\bm x}_0)/\beta_t$
\STATE Update $\bm{x}$ by \eqref{adjusted_ddpm}, then $t\leftarrow t^\textrm{next}$
\ENDWHILE
\STATE Output the generated sample $\bm{x}$
\end{algorithmic}
\gthchangeend
\end{algorithm}

\subsection{Per-Dimension Encoding as a Fixed-Readout Alternative}\label{sec:8q}
The absorption mechanism derived above is a statement about the reverse process, not about the probability encoding, and it is instructive to instantiate it for a different encoding. We do so for the angle encoding of Section~\ref{sec:enc}, in which every coordinate is carried by its own qubit. The resulting scheme is of independent interest because its readout contains no trainable parameters at all, so that whatever denoising takes place must take place inside the circuit.

A sample $\bm{x}\in[-A,A]^N$ is loaded by rotating qubit $i$ so that its Pauli-$Z$ expectation reproduces the corresponding coordinate,
\begin{equation}
\langle Z_i\rangle=\frac{x_i}{A},\qquad i=1,\dots,N,\label{angleenc}
\end{equation}
which is realized by an $R_Y$ rotation of angle $\arccos(x_i/A)$, and the clean signal is read back by inverting the same map,
\begin{equation}
\hat{x}_{0,i}=A\,\langle \hat{Z}_i\rangle.\label{readout8q}
\end{equation}
The readout \eqref{readout8q} is thus dictated by the encoding and carries no free parameters, in contrast with the trainable map \eqref{eq:readout}. Its price is the qubit count, $n_q=N$ instead of $n_q=\log_2 N$, so the scheme is affordable only for small $N$, for instance after a code has reduced the data dimension. The data also no longer needs to be zero-sum, since separate qubits carry the coordinates and no normalization couples them. The forward process \eqref{diff} uses an isotropic Gaussian and $\mathbf{\Xi}$ is replaced by $\mathbf{I}$ throughout.

\gthchanged{Measuring qubit $i$ in the computational basis returns a $\pm1$ outcome whose mean is $\langle Z_i\rangle$, i.e.\ a Bernoulli variable of variance $1-\langle Z_i\rangle^2$, and distinct qubits are measured independently. With $M_i$ shots on coordinate $i$, the CLT gives $\hat{Z}_i-\langle Z_i\rangle\mathrel{\overset{\mathrm{approx.}}{\sim}}\mathcal{N}\big(0,(1-\langle Z_i\rangle^2)/M_i\big)$, so that by \eqref{readout8q} under the stated independent-measurement assumption, the readout error is approximately Gaussian with a diagonal covariance,}
\gthchangebegin
\begin{equation}
\begin{aligned}
\hat{\bm{x}}_0=\hat{\bm{x}}_0^{\star}+\bm{\delta},\qquad
&\bm{\delta}\mathrel{\overset{\mathrm{approx.}}{\sim}}\mathcal{N}\big(\bm{0},\mathbf{D}_t\big),\\
&[\mathbf{D}_t]_{ii}=\frac{A^2\big(1-\langle Z_i\rangle^2\big)}{M_i}.
\end{aligned}\label{delta8q}
\end{equation}
\gthchangeend
Equation \eqref{delta8q} is the counterpart of Theorem~\ref{the:proberr} for the present encoding, and it is markedly simpler. The multinomial coupling $-p_ip_j$ of \eqref{sigma} is absent, because the coordinates are no longer competing outcomes of one and the same measurement.

The shot budget of the eight-qubit model follows by reusing the general derivation. Substituting \eqref{delta8q} into the reverse step \eqref{denoising} and repeating the argument of Section~\ref{sec:algo} verbatim, the covariance contract becomes
\begin{equation}
\sigma_t^2\mathbf{I}-k_t^2\,\mathbf{D}_t\succeq0.\label{requirement8q}
\end{equation}
Both matrices are diagonal, so \eqref{requirement8q} decouples into $N$ scalar inequalities, one per coordinate, and the measurements can be budgeted per dimension,
\begin{equation}
M_{t,i}=\operatorname{clamp}_{[M_\mathrm{min},M_\mathrm{max}]}\left(\left\lceil\frac{k_t^2}{\sigma_t^2}\,A^2\big(1-\langle Z_i\rangle^2\big)\right\rceil\right).\label{M8q}
\end{equation}
The residual noise to be injected is then diagonal as well, with $i$-th variance $\sigma_t^2-k_t^2A^2(1-\langle Z_i\rangle^2)/M_{t,i}$, so no eigendecomposition is needed and the residual-noise step of Algorithm~\ref{alg:ddpm} collapses into $N$ independent scalar corrections.

Two features of \eqref{M8q} are worth noting. First, the budget is largest where $|\langle Z_i\rangle|$ is small, that is, where the Bernoulli variance of a coordinate is maximal, and smallest for coordinates already close to $\pm A$. The schedule therefore adapts across coordinates as well as along the trajectory, which the shared budget \eqref{M} cannot do. Second, the mechanism itself is untouched, since $\sigma_t$ enters \eqref{M8q} exactly as it enters \eqref{M}, so that Corollary~\ref{cor:eta} applies verbatim. The two encodings differ only in their resource profile, a logarithmic qubit count with a coupled, trainable readout, against a linear one with a decoupled, parameter-free readout.

\gthchanged{\section{Mechanism Illustration and Validation on a Gaussian Mixture}}
We begin with a Gaussian mixture. The data lives on the zero-sum hyperplane, so the task is a degenerate seven-dimensional distribution embedded in $\mathbb{R}^{8}$. A mixture of this kind is easy to visualize and its ground truth is known exactly, which is what we need in order to check that the under-sampled quantum sampler still recovers several modes instead of collapsing onto one. It also lets us take the predictions of Section~\ref{sec:algo} one at a time, namely the shape of the shot budget \eqref{M}, the two bounds that clamp it, the cost of determinism of Corollary~\ref{cor:eta}, what the trainable readout contributes, and the per-dimension variant of Section~\ref{sec:8q}.

\vspace{0.6\baselineskip}

\subsection{Experimental Setup}
We consider a degenerate Gaussian mixture whose $K=N-1=7$ free coordinates are lifted onto the zero-sum hyperplane of $\mathbb{R}^{N}$ with $N=8$ through \eqref{lift}. The three components are equally weighted and their centers lie on that hyperplane, $\bm{1}^\top\bm{c}_k=0$, with
\begin{align*}
\bm{c}_1&=(0.55,0.40,0.20,0.05,-0.05,-0.20,-0.40,-0.55),\\
\bm{c}_2&=(0.40,0.55,-0.55,-0.40,-0.20,-0.05,0.05,0.20),\\
\bm{c}_3&=(0.45,-0.45,0.35,-0.35,0.20,-0.20,0.05,-0.05).
\end{align*}
Each component is isotropic on the hyperplane with standard deviation $\sigma_0=0.12$, so the mixture density reads
\begin{equation}
p(\bm{x})=\frac{1}{3}\sum_{k=1}^{3}\mathcal{N}\!\left(\bm{x};\,\bm{c}_k,\,\sigma_0^2\mathbf{\Xi}\right),\label{eq:gmm}
\end{equation}
where $\mathbf{\Xi}$ confines the covariance to the hyperplane. We draw a sample by picking a component $k$ uniformly and setting $\bm{x}_0=\bm{c}_k+\sigma_0\mathbf{\Xi}\bm{z}$ with $\bm{z}\sim\mathcal{N}(\bm{0},\mathbf{I})$, which gives $\bm{1}^\top\bm{x}_0=0$ by construction and leaves nothing to project. The data sits in the hyperplane box $\mathcal{S}^{N-1}(A)$ of \eqref{restriction} with bound $A=3$.

The denoiser is a data re-uploading circuit on $n_q=3$ qubits, one computational-basis probability per entry. At each reverse step the current iterate $\bm{x}$ is mapped to a probability vector $\bm{q}$ by \eqref{forwardmap} and loaded through the hybrid amplitude-phase embedding $\ket{q(\tau)}=\sum_i\sqrt{q_i}\,e^{jf_i(\tau)}\ket{i}$, with $\tau\in[0,1]$ the normalized diffusion step. The phase is the normalized time embedding \eqref{timeemb}--\eqref{cnorm}. We fix its shape parameters at $a=0.6$ and $b=0.15$ and leave them untuned, and cap the largest excursion at $\phi_{\max}=\pi$ so that every phase stays within half a revolution. The cap matters. Left unnormalized, the profile runs past a full revolution and aliases the diffusion step, and at $T=1000$ the embedding would no longer separate neighbouring steps. The circuit then applies $L=12$ re-uploading layers. Each layer rotates every qubit by $R_Y$ and $R_Z$ with angles linear in the re-uploading feature $[\,1,\;1.4(N\bm{q}-\bm{1}),\;\tau\,]$, followed by a ring of CNOT gates. Measuring in the computational basis returns the estimates $\tilde{\bm{p}}$.

For the readout we take the trainable linear readout \eqref{eq:readout}, initialized at the fixed affine decoding \eqref{backmap} with $\mathbf{W}=AN\mathbf{I}=24\mathbf{I}$ and $\bm{b}=-A\bm{1}$ and trained jointly with the circuit parameters $\bm{\theta}$. We constrain $\mathbf{W}$ to be diagonal, which keeps the readout parameter count linear in $N$.

Training uses the $\bm{x}_0$-prediction target of \eqref{target} at $T=1000$ diffusion steps under the schedule \eqref{eq:schedule} with $\beta_{\min}=0.10$ and $\beta_{\max}=0.95$, over $12000$ training pairs and $5\times10^{4}$ Adam iterations with a learning rate decaying from $0.02$. Generation uses the sampler \eqref{denoising} at $\eta=1$ with $20$ reverse steps unless we say otherwise. The per-step shot count comes online from \eqref{M}, scaled by a safety factor of $1.10$ and clamped to $[M_\mathrm{min},M_\mathrm{max}]=[2,\,2\times10^{5}]$. Each reported figure uses $3000$ generated samples, and each ablation is repeated over five independent runs, from which we report the mean and the standard deviation.

The mixture is well separated, so we can report something directly interpretable, the resolved fraction. Let $\{\bm{x}^{(i)}\}_{i=1}^{n}$ be the generated samples, $\{\bm{c}_k\}_{k=1}^{3}$ the centers, and $d_{\min}=\min_{k\neq l}\|\bm{c}_k-\bm{c}_l\|_2$ the minimum inter-center distance. The resolved fraction is
\begin{equation}
\mathrm{rf}=\frac{1}{n}\sum_{i=1}^{n}\mathbbm{1}\!\left[\min_{k}\|\bm{x}^{(i)}-\bm{c}_k\|_2<\tfrac{1}{2}d_{\min}\right],\label{eq:rf}
\end{equation}
\gthchanged{the fraction of generated points within half the minimum inter-center distance of a true center. It measures proximity to the centers; we assess mode coverage alongside the distribution plots and $\mathrm{MMD}^2$. The latter is the squared maximum mean discrepancy between generated samples $\{\bm{x}^{(i)}\}_{i=1}^{n}$ and target samples $\{\bm{y}^{(j)}\}_{j=1}^{m}$,}
\begin{equation}
\begin{split}
\mathrm{MMD}^2={}&\frac{1}{n^2}\sum_{i,i'}k(\bm{x}^{(i)},\bm{x}^{(i')})+\frac{1}{m^2}\sum_{j,j'}k(\bm{y}^{(j)},\bm{y}^{(j')})\\
&-\frac{2}{nm}\sum_{i,j}k(\bm{x}^{(i)},\bm{y}^{(j)}),
\end{split}\label{eq:mmd}
\end{equation}
with the Gaussian kernel $k(\bm{x},\bm{y})=\exp(-\gamma\|\bm{x}-\bm{y}\|_2^2)$, whose bandwidth $\gamma$ is the inverse of the median pairwise squared distance. Measurement cost is reported as shots per generated sample, summed along the reverse trajectory.

\subsection{Generated Distribution and Denoising Trajectory}
\begin{figure}[t]
\centering
\includegraphics[width=\columnwidth]{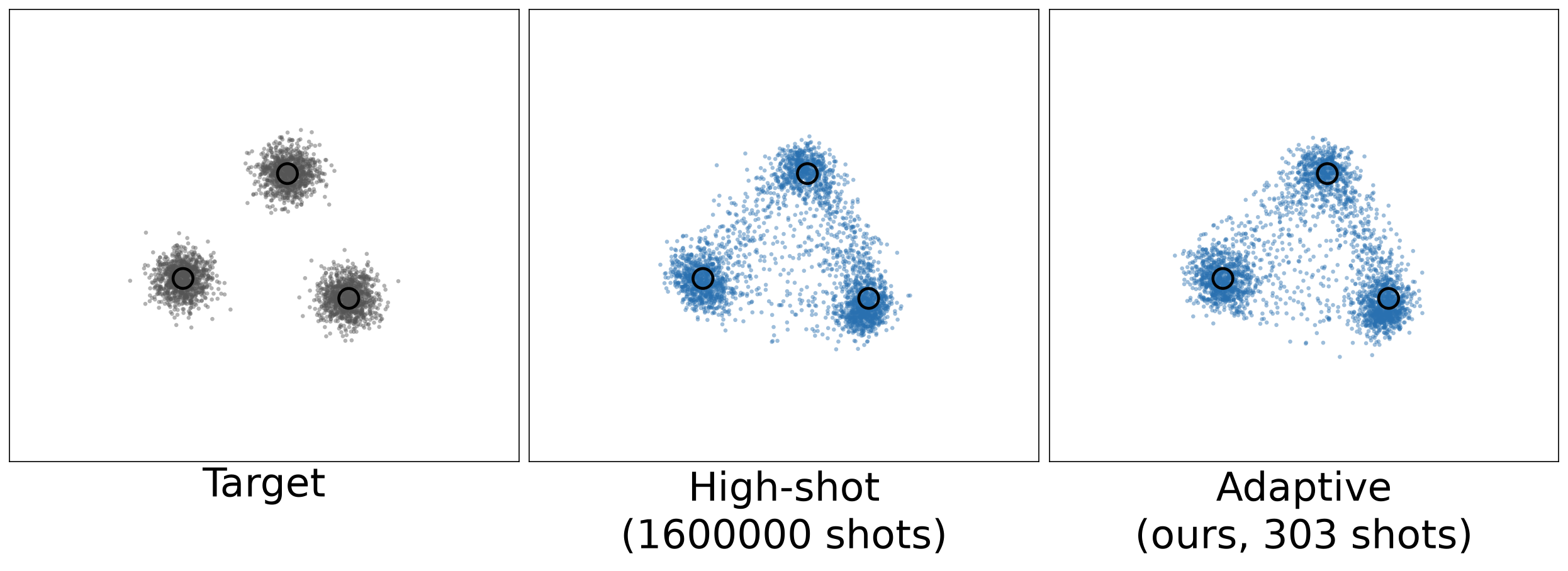}
\caption{\gthchanged{Target, high-shot reference, and adaptive sampler on the three-center plane.}}
\label{fig:gmm_triptych}
\end{figure}

\gthchanged{Fig.~\ref{fig:gmm_triptych} places the target mixture next to what the quantum sampler produces, first with a high-shot reference readout and then at the adaptive operating point. The two are hard to tell apart. On both metrics the adaptive run delivers comparable quality, at $\mathrm{rf}=0.941$, while spending $303$ shots per generated sample, about $5300\times$ fewer measurements than the reference. The qualitative point comes first: the three clusters are recovered rather than merged, and until that holds no efficiency claim means anything.}

\begin{figure*}[t]
\centering
\includegraphics[width=0.70\textwidth]{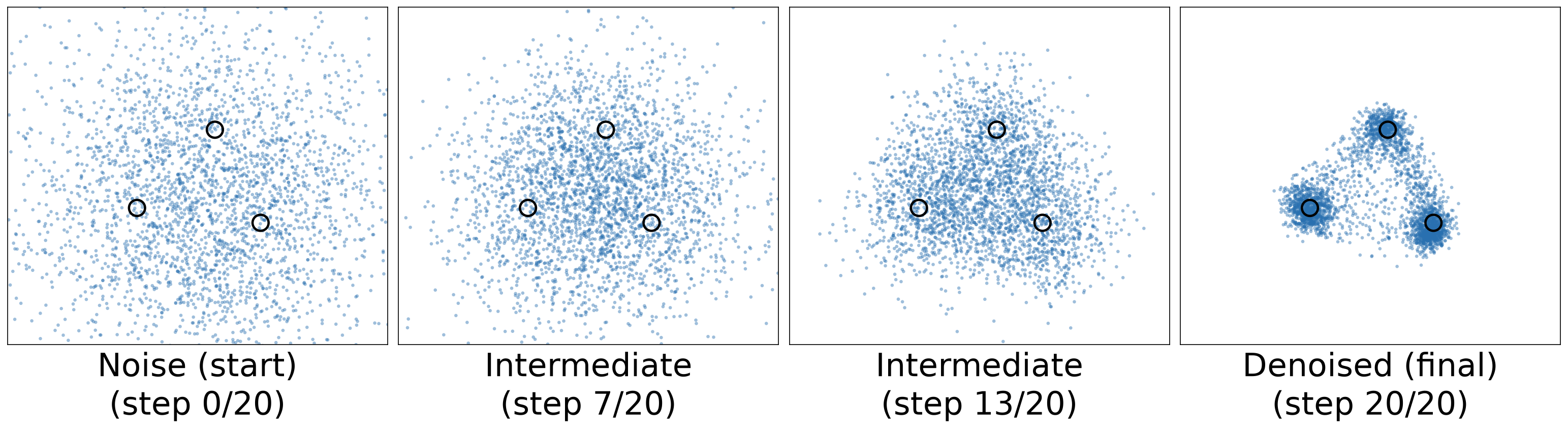}
\caption{Reverse-diffusion procedure of the under-sampled quantum sampler.}
\label{fig:gmm_denoise}
\end{figure*}

Fig.~\ref{fig:gmm_denoise} follows the reverse trajectory at the twenty reverse steps we use as our default operating point. The sampler starts from an isotropic cloud on the hyperplane and the three modes pull apart only over the last few steps, once the injected noise $\sigma_t$ has become small and the denoiser has to be accurate. That asymmetry along the trajectory is what \eqref{M} feeds on, and it is the subject of the next subsection.

\subsection{The Adaptive Shot Budget}\label{sec:gmm_budget}
\begin{figure}[t]
\centering
\includegraphics[width=0.7\columnwidth]{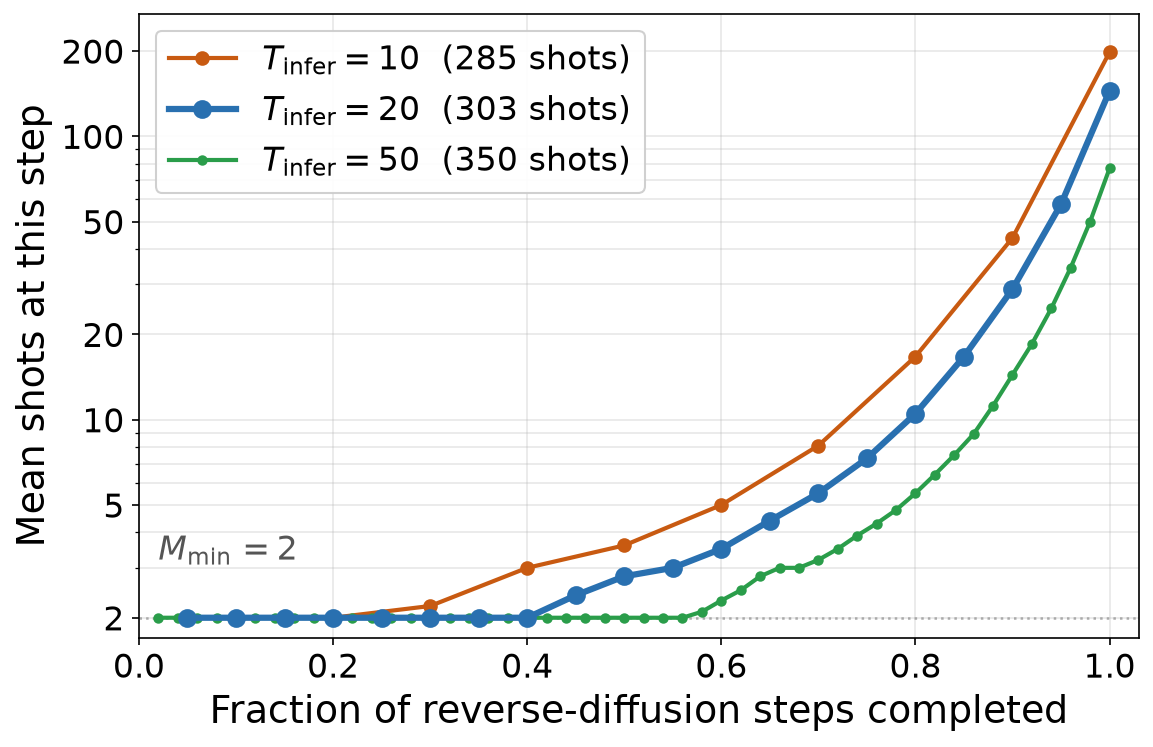}
\caption{Shots per reverse step under the adaptive budget \eqref{M}.}
\label{fig:gmm_shots}
\end{figure}

The budget \eqref{M} is recomputed at every reverse step from the current state, and Fig.~\ref{fig:gmm_shots} shows the profile it produces for reverse processes of ten, twenty, and fifty steps. All three trace the same shape. The demand sits on the lower bound $M_\mathrm{min}$ over the early part of the trajectory, where the sampler measures as little as the clamp allows, and then climbs steeply toward the end, with more than half of the total spent on the last two steps alone. Section~\ref{sec:budget} predicts this. The injected noise $\sigma_t$ is large early on and can absorb a very coarse readout, and it shrinks as the trajectory proceeds, where the readout has to become accurate. The three schedules also spend a similar total, so the profile is set by the noise schedule rather than by the number of reverse steps.

\begin{figure}[t]
\centering
\includegraphics[width=0.65\columnwidth]{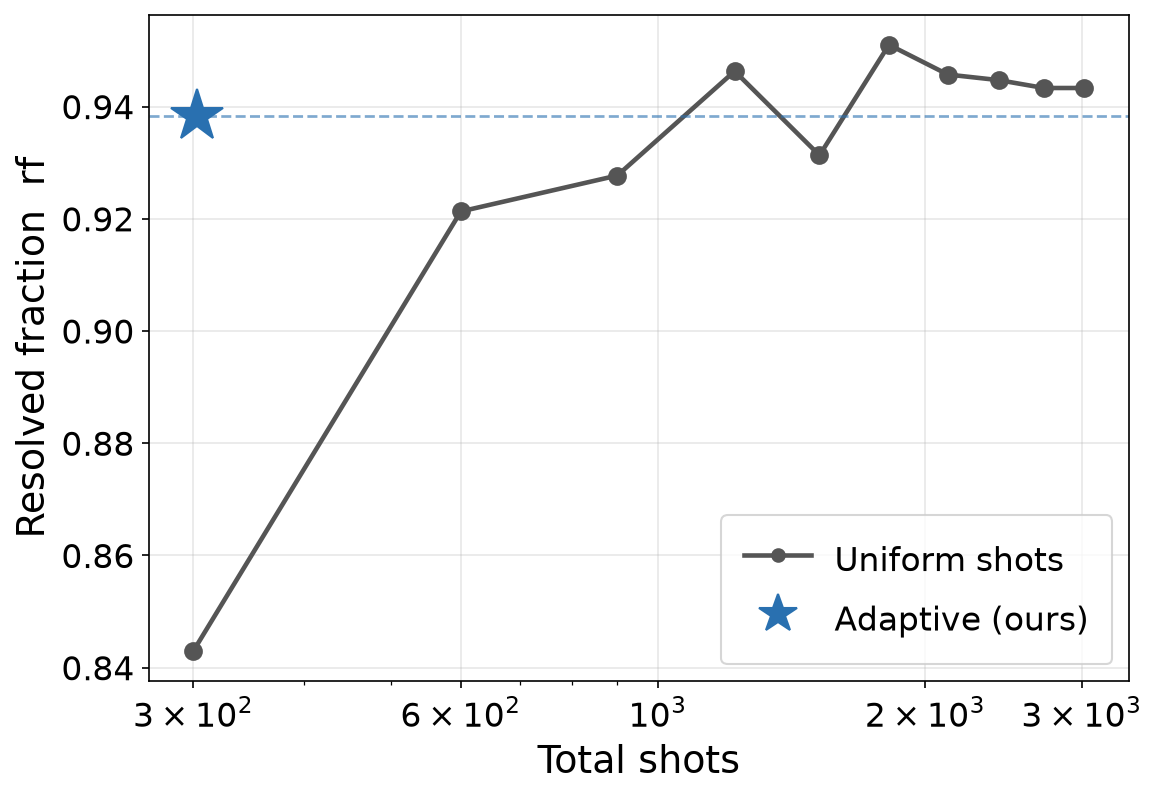}
\caption{Adaptive versus uniform shot allocation.}
\label{fig:gmm_uniform}
\end{figure}

The same total budget can buy very different quality depending on how it is spread. Fig.~\ref{fig:gmm_uniform} makes that comparison against a uniform allocation over the same twenty steps. At a matched budget of a few hundred shots per sample the adaptive schedule is well ahead, and the uniform one needs about four times as many measurements before it catches up on the resolved fraction. Spreading the measurements evenly wastes them on steps that do not need them and starves the steps that do. The gap is widest in the low-budget regime, which is the regime that matters for quantum sampling.

\subsection{The Bounds on the Budget}\label{sec:gmm_bounds}
\begin{figure}[t]
\centering
\includegraphics[width=\columnwidth]{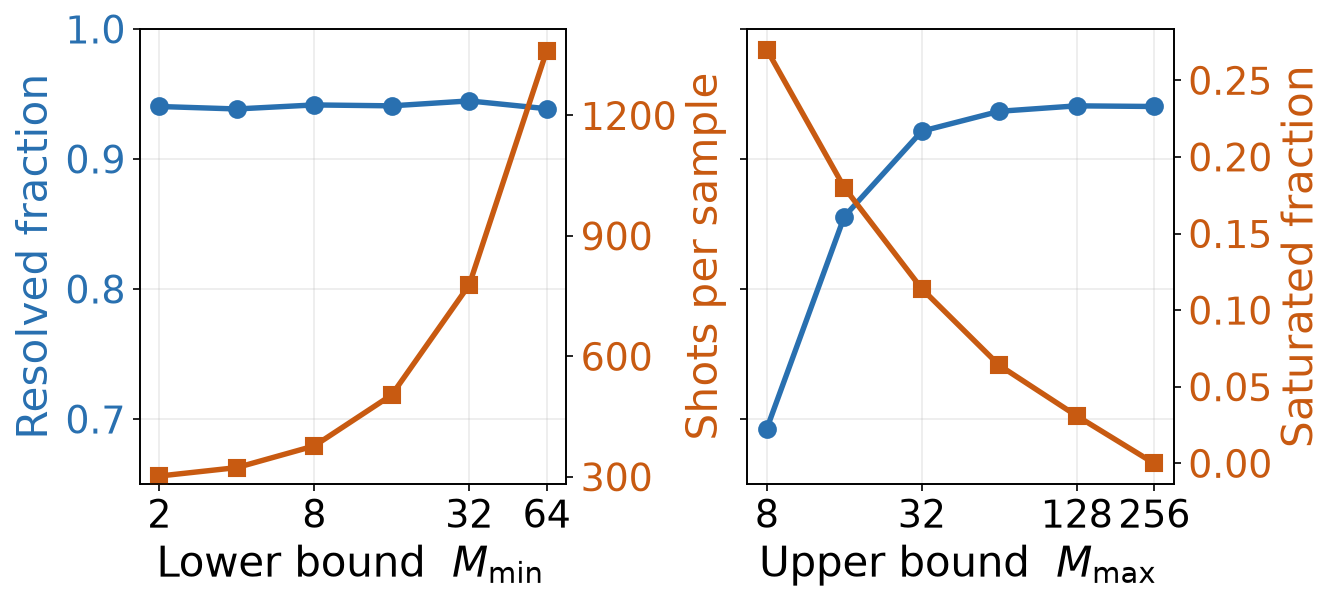}
\caption{Sensitivity of the budget to its two bounds, over five runs.}
\label{fig:gmm_bounds}
\end{figure}

The budget \eqref{M} has only two free parameters and they come from different places. The lower bound $M_\mathrm{min}$ is a modelling safeguard. Theorem~\ref{the:proberr} is a central-limit statement, and the usual rule of thumb for the multinomial CLT asks for $M p_i\gtrsim5$, which at $p_i\approx1/N$ would demand $M\gtrsim 5N=40$. The upper bound $M_\mathrm{max}$ is a resource limit set by what measurement costs. Fig.~\ref{fig:gmm_bounds} sweeps them one at a time, the lower bound on the left and the upper bound on the right.

\gthchanged{Raising the lower shot bound from 2 to 64 leaves the resolved fraction essentially unchanged and $\mathrm{MMD}^2$ unchanged within its standard deviation, while increasing cost by about a factor of four and a half. The two-shot steps occur mainly early in the trajectory, where the noise budget is large. We therefore select $M_\mathrm{min}=2$ empirically, despite operating outside the Gaussian-approximation regime.}

\gthchanged{The upper bound behaves in the opposite way. It truncates \eqref{M} where the budget is largest, at the final steps, so a cap below the peak demand leaves the PSD condition \eqref{requirement} unsatisfied. The residual covariance then acquires negative eigenvalues, which are clipped to zero. For the true-covariance model, this removes external noise in the affected directions, but cannot remove measurement noise that already exceeds the target. The resulting failure is a mismatch in total noise covariance, not a shortage of injected randomness. The effect is severe. As $M_\mathrm{max}$ rises from below the peak demand, the resolved fraction climbs steeply while the saturated fraction falls, and once the cap clears the peak the saturated fraction reaches zero and the resolved fraction settles at the value of the unclamped schedule. Quality tracks the saturated fraction rather than the total budget, which is the signature of a broken covariance contract and not of a shortage of measurements.}

\gthchanged{We use $M_\mathrm{min}=2$ and $M_\mathrm{max}=2\times10^5$: the lower bound saves measurements in early steps, while the upper bound remains above peak demand and never becomes active.}

\subsection{Injected Noise and the Cost of Determinism}\label{sec:gmm_eta}
\begin{figure}[t]
\centering
\includegraphics[width=0.65\columnwidth]{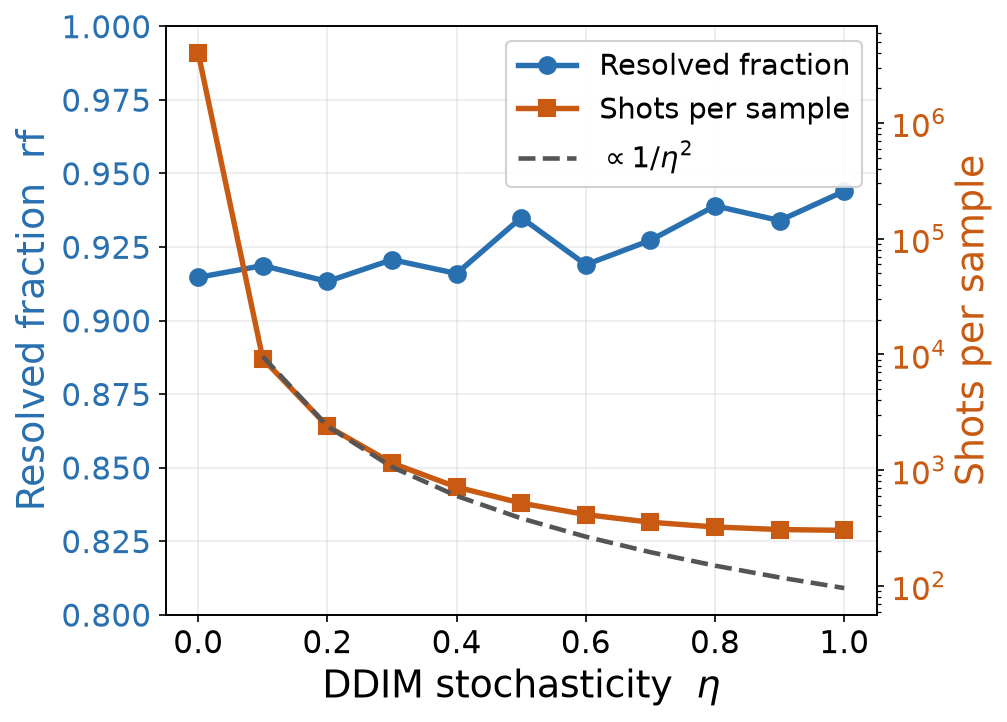}
\caption{Effect of the DDIM parameter $\eta$.}
\label{fig:gmm_eta}
\end{figure}

The parameter $\eta$ picks a member of the DDIM family, from the deterministic sampler at $\eta=0$ to the stochastic DDPM sampler at $\eta=1$, and the framework of Section~\ref{sec:algo} covers all of them. Generation quality bears this out. Across the sweep of Fig.~\ref{fig:gmm_eta} the resolved fraction is essentially unchanged, so any member of the family may be used, DDIM and DDPM alike.

\gthchanged{The measurement cost is another matter, and Corollary~\ref{cor:eta} says how it should behave. A sampler made more deterministic has to be measured more, at a rate of $1/\eta^{2}$. The sweep follows that law closely wherever the budget sits well above its lower bound, where halving $\eta$ costs about a factor of $3.8$, close to the predicted $4$. As $\eta$ grows the budget falls onto $M_\mathrm{min}$ over the early steps, the clamp takes over, and the growth flattens toward the stochastic end. The deterministic limit is the extreme case. At zero stochasticity, the prescribed update has no injected-noise budget for compensation. We therefore use the capped high-shot setting in this comparison; its cost is four orders of magnitude above that of the stochastic sampler. Any remaining finite-shot error is an uncompensated approximation to the deterministic update.}

\gthchanged{Across this sweep, generation quality remains approximately flat while measurement cost falls toward the stochastic end. We therefore use $\eta=1$, the least expensive setting in the comparison.}

\subsection{The Role of the Readout}\label{sec:gmm_readout}\label{sec:gmm_8q}
Section~\ref{sec:absorb} makes a specific claim about the linear readout, that it reshapes the covariance of the finite-shot noise into $\mathbf{\Xi}\mathbf{W}^\top\mathbf{\Sigma}\mathbf{W}\mathbf{\Xi}$ rather than simply rescaling it, and so changes how many shots the sampler needs. The upper block of Table~\ref{tab:gmm_readout} tests this inside the logarithmic encoding, with the frozen affine decoding of Corollary~\ref{cor:fixed} on one side and our default diagonal readout, which carries $N$ parameters, on the other.

The frozen readout loses on both axes. Its resolved fraction is markedly lower and it demands roughly thirty times as many shots. Training the readout therefore improves the fit and cuts the measurement cost at the same time, which is the behaviour \eqref{M} would lead one to expect. A readout that steers the shaped covariance into directions the diffusion noise can absorb lowers its largest eigenvalue, and the budget follows. It does so without growing large. The analytic scale of the readout is $AN$, while the learned diagonal readout settles at a spectral norm some seven times smaller, because the circuit learns to output probabilities further from uniform, so a smaller gain already covers the data range.

\gthchanged{This comparison does not by itself isolate how much denoising is supplied by the trainable readout. The fixed affine baseline uses $\mathbf W=AN\mathbf I_8=24\mathbf I_8$ and $\bm b=-A\bm1$, with $A=3$ and $N=8$, where $\mathbf I_8$ is the eight-dimensional identity matrix. The factor $AN$ reverses the scaling of \eqref{forwardmap} such that probability deviations must be converted back to data units. In the unclipped regime, the baseline requires the circuit to produce the prescribed probability target accurately, which is demanding for twelve layers on three qubits. The per-dimension encoding of Section~\ref{sec:8q} settles it. Its readout \eqref{readout8q} is the exact inverse of its own encoding rather than something imposed from outside, so it carries no trainable parameters and nothing has to be matched, and whatever denoising happens has to happen inside the circuit. We run it on the same mixture with $n_q=N=8$ qubits, the same diffusion schedule, and the per-dimension budget \eqref{M8q}.}

\begin{figure*}[t]
\centering
\includegraphics[width=0.70\textwidth]{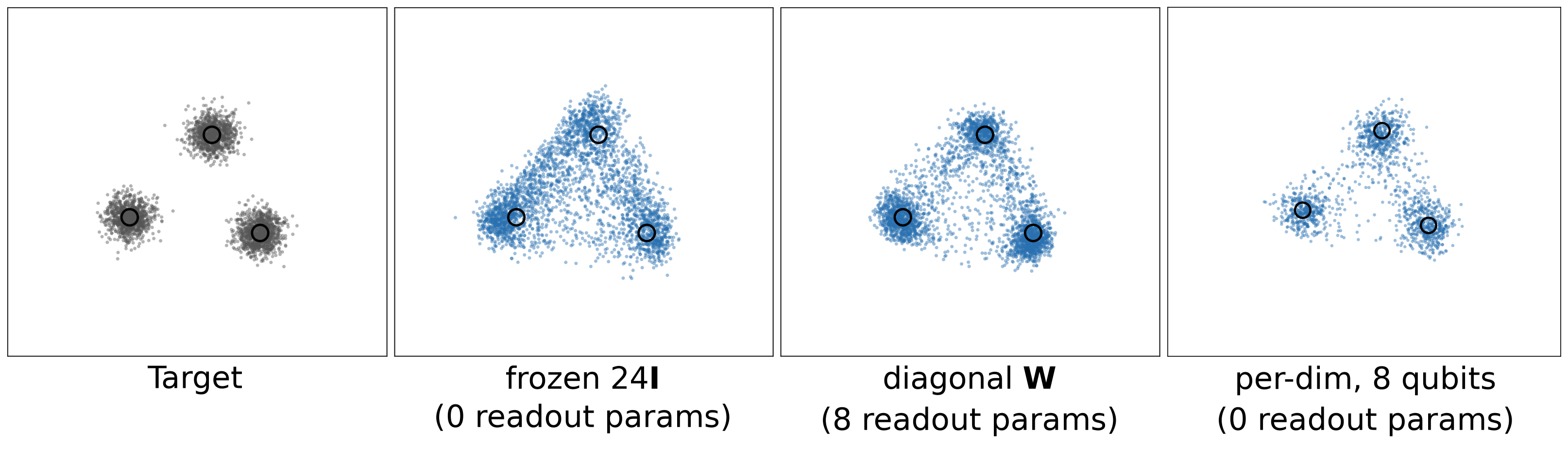}
\caption{\gthchanged{Generated distributions under the readouts in Table~\ref{tab:gmm_readout}, where ``frozen $24\mathbf I$'' denotes the fixed affine baseline.}}
\label{fig:gmm_readout}
\end{figure*}

The lower block of Table~\ref{tab:gmm_readout} and the last panel of Fig.~\ref{fig:gmm_readout} report what comes out. The adaptive sampler again matches the shot-noise-free readout, here at about $1600\times$ fewer measurements. The comparison with the frozen row is the point. With no trainable readout parameters at all the per-dimension encoding reaches the same quality, whereas the same absence of parameters noticeably lowers the resolved fraction of the logarithmic encoding. The structural learning is thus done by the quantum circuit, and the trainable readout of the logarithmic encoding improves the fit and the measurement cost without being the agent of the denoising. The absorption mechanism and the adaptive schedule carry over unchanged to an encoding whose finite-shot covariance is diagonal rather than multinomial, as Section~\ref{sec:8q} predicts. The price is the qubit count, which grows from $\log_2 N$ to $N$, and that is what keeps the per-dimension encoding to low-dimensional data until a code reduces the dimension, as in Section~\ref{sec:img_perdim}.

\begin{table}[t]
\centering
\caption{Readout comparison, at $20$ reverse steps and $\eta=1$.}
\label{tab:gmm_readout}
\begin{tabular}{lcccc}
\toprule
readout & params & $\mathrm{rf}$ & $\mathrm{MMD}^2$ & shots \\
\midrule
\multicolumn{5}{l}{Logarithmic encoding, $n_q=\log_2 N=3$}\\
\gthchanged{Fixed affine} & $0$ & $0.856$ & $0.0162$ & $9173$ \\
diagonal & $N=8$ & $0.941$ & $0.0049$ & $\mathbf{303}$ \\
\midrule
\multicolumn{5}{l}{Per-dimension encoding, $n_q=N=8$}\\
fixed $A\langle Z\rangle$ & $0$ & $\mathbf{0.945}$ & $\mathbf{0.0033}$ & $991$ \\
\bottomrule
\end{tabular}
\end{table}

\subsection{Ablation on the Encoding and the Time Embedding}\label{sec:ablation}
We close by isolating the encoding of Section~\ref{sec:enc}. Table~\ref{tab:gmm_encoding} crosses the three encodings with the presence or absence of the time embedding, five runs each, everything else held fixed. Switching the time embedding off removes the azimuthal phase, at which point the hybrid encoding coincides with probability encoding by construction, and the two rows agree.

The time embedding is not optional. Removing it raises the $\mathrm{MMD}^2$ by a factor of $3\sim 4$ for every encoding and drops the resolved fraction below $0.89$. A denoiser that cannot tell which diffusion step it is on cannot invert the forward process, and that is the dominant effect in the table.

With the time embedding present, the hybrid encoding is the best on both metrics, ahead of probability encoding and amplitude encoding. The margin over probability encoding is what writing the diffusion step into the free phases is worth, and it costs no qubits. The margin over amplitude encoding is smaller, and amplitude encoding is in any case not usable here for the structural reason given in Section~\ref{sec:enc}. Its output does not live in the space of its input, so the denoiser cannot be iterated without an extra reconstruction step, and we report the comparison only for completeness.

\begin{table}[t]
\centering
\caption{Encoding and time embedding, at $20$ reverse steps, $\eta=1$.}
\label{tab:gmm_encoding}
\begin{tabular}{llcc}
\toprule
encoding & time & $\mathrm{rf}$ & $\mathrm{MMD}^2$ \\
\midrule
amplitude   & on  & $0.934\pm0.005$ & $0.0069\pm0.0009$ \\
probability & on  & $0.930\pm0.003$ & $0.0069\pm0.0012$ \\
hybrid      & on  & $\mathbf{0.939\pm0.005}$ & $\mathbf{0.0062\pm0.0006}$ \\
\midrule
amplitude   & off & $0.885\pm0.003$ & $0.0207\pm0.0006$ \\
probability & off & $0.878\pm0.007$ & $0.0246\pm0.0007$ \\
hybrid      & off & $0.878\pm0.007$ & $0.0246\pm0.0007$ \\
\bottomrule
\end{tabular}
\end{table}

\begin{figure*}[t]
\centering
\includegraphics[width=\textwidth]{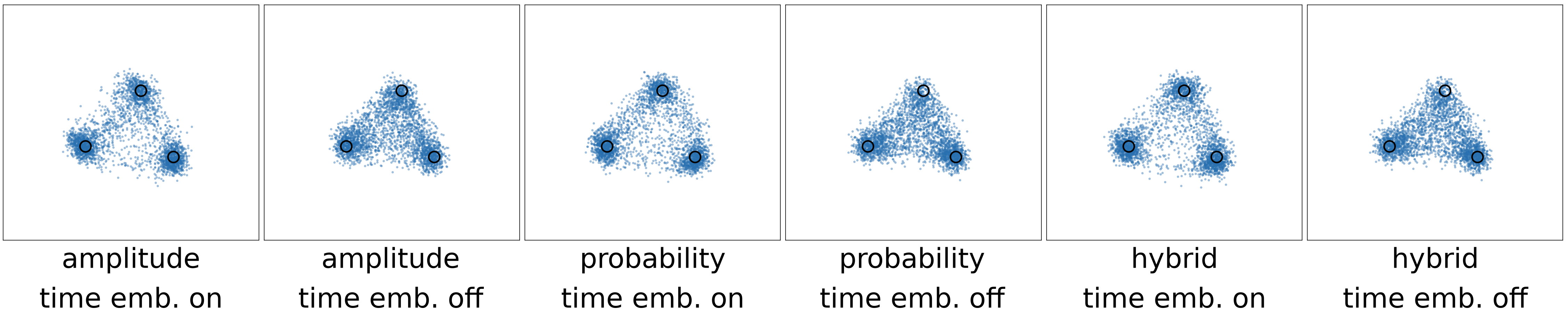}
\caption{Generated distributions for the six cells of Table~\ref{tab:gmm_encoding}.}
\label{fig:gmm_encoding}
\end{figure*}

\section{Image Generation Examples}
\gthchanged{We next apply the method to MNIST in a learned latent space. To keep the quantum circuit small, we compress each $28\times28=784$-pixel image into a low-dimensional latent. A fixed variational autoencoder (VAE) supplies this representation and the image decoder. The QNN predicts the clean latent at each reverse step; the full loop starts from random latent noise and generates a new sample, which the VAE decodes into an image. We test the trainable-diagonal and fixed per-dimension readouts, then increase the latent dimension.}

\gthchanged{\subsection{VAE-Based Compression}\label{sec:img_code}}
We compress each image into a low-dimensional latent with a variational autoencoder (VAE)~\cite{kingma2014vae}. It produces the $K=7$ latent coordinates that we lift onto the eight-dimensional zero-sum hyperplane, then rescale and clip as in Section~\ref{sec:diff}.

\begin{figure}[t]
\centering
\includegraphics[width=0.9\columnwidth]{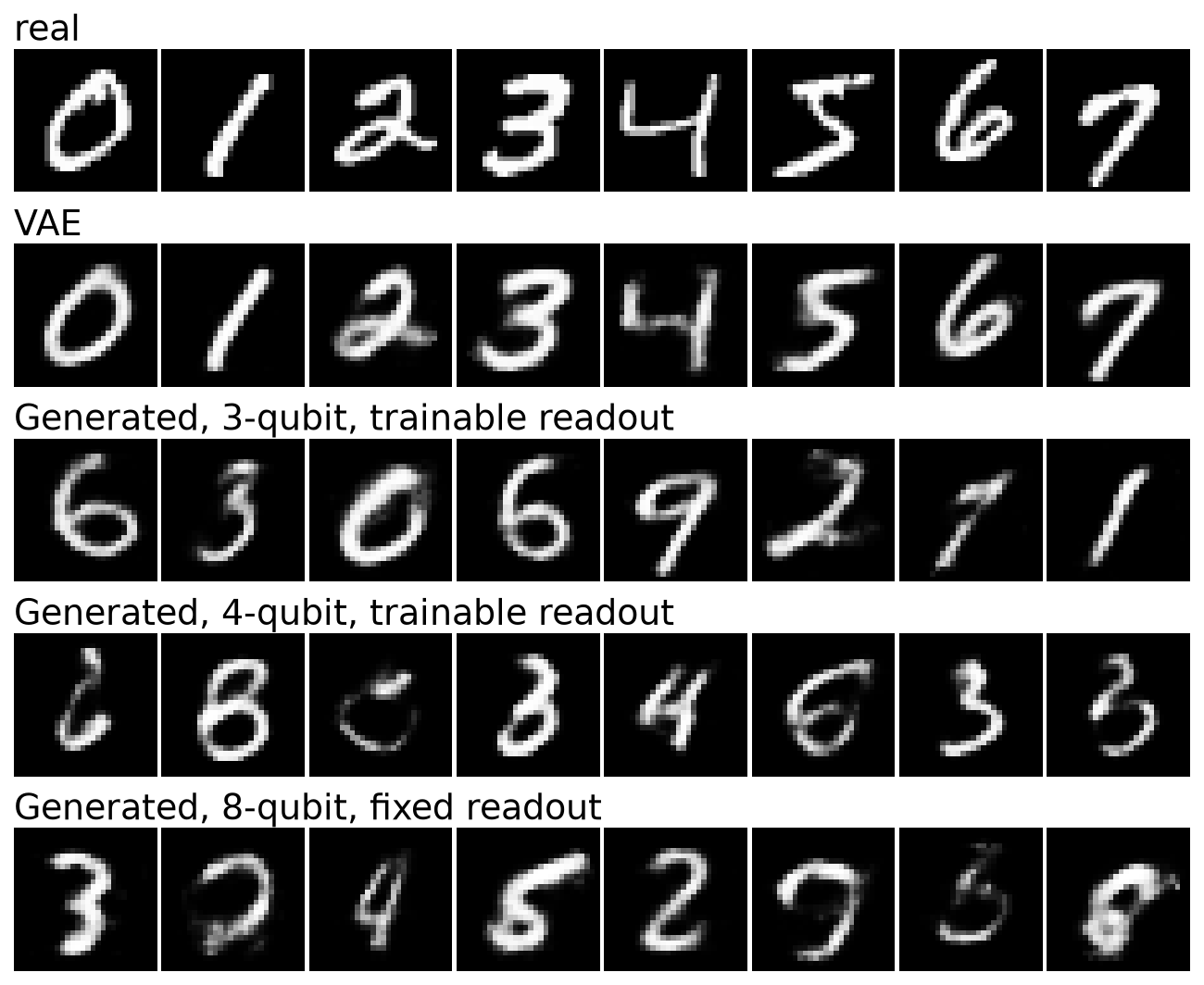}
\caption{\gthchanged{MNIST examples: real images and VAE reconstructions (top two rows), followed by QNN-generated samples from latent diffusion, decoded by the VAE.}}
\label{fig:img}
\end{figure}

The round trip through the VAE keeps the digits intact, and Fig.~\ref{fig:img} shows this before the sampler is run at all. The real row and the VAE row carry the same legible digits, and a classifier that scores $97.1\%$ on the originals still reaches $94.4\%$ on the reconstructions. Working in this latent therefore ties the quantum resource to the latent dimension rather than to the image resolution, and we use it for the remainder of this section.

\subsection{Latent-Domain Quantum DDPM}\label{sec:img_latent}
\gthchanged{The QNN retains the same role as in the Gaussian-mixture experiment: it predicts the clean state from the current noisy iterate and the diffusion time. Here, the data coordinates come from the fixed VAE encoder. Its seven latent coordinates are lifted onto the eight-dimensional zero-sum hyperplane and processed by the same hybrid probability-and-time-phase encoding, a three-qubit circuit with 20 re-uploading layers, and a trainable diagonal readout. We train the denoiser to predict the clean latent state over 1,000 diffusion steps and use the same shot-adaptive reverse sampler for generation. Each generated trajectory begins with random noise. The final state is mapped back to the seven VAE coordinates and decoded into a 28-by-28 image, without post-processing.}

\gthchanged{The three-qubit row of Fig.~\ref{fig:img} shows generated samples covering all ten digit classes, with $49.1\%$ receiving classifier confidence above 0.9. The classifier is used only for evaluation.}

The efficiency result of Section~\ref{sec:algo} carries over to real images unchanged. The adaptive schedule reaches $\mathrm{MMD}^2=0.0312$ in image space with $430$ shots per generated sample, about $3700\times$ fewer measurements than the shot-noise-free readout at matched quality. It also survives a larger latent. The four-qubit row of Fig.~\ref{fig:img} grows the latent to $K=15$, and the digits stay legible at $\mathrm{MMD}^2=0.0259$ with a saving of about $2400\times$, so the mechanism is not tied to a single latent dimension.

\subsection{Per-Dimension Encoding in the Latent Domain}\label{sec:img_perdim}
The per-dimension encoding of Section~\ref{sec:8q} offers an alternative to the logarithmic probability encoding. Coordinate $i$ is carried by qubit $i$ and read back through the fixed map \eqref{readout8q}, so the readout holds no trainable parameters and whatever denoising happens has to happen inside the circuit. Its cost is a qubit count linear in the data dimension, which rules it out in pixel space, since a $784$-pixel image would call for $784$ qubits. That is why the encoding has so far been confined to low-dimensional synthetic data. The latent formulation removes the obstacle. With only $8$ dimensions to carry, we can run both encodings on the same latent, under the same diffusion schedule and the same under-sampling rule.

The eight-qubit row of Fig.~\ref{fig:img} shows the result. With $8$ qubits and a fixed readout, the model reaches $\mathrm{MMD}^2=0.0143$ in image space using $993$ shots per sample, about $1600\times$ fewer than the shot-noise-free readout, and $55.1\%$ of its samples are classified confidently. It is the best of the three configurations. The readout carries no free parameters, so the denoising is the work of the quantum circuit alone. That this configuration matches the trainable-readout model, and here slightly exceeds it, reproduces on images what Section~\ref{sec:gmm_8q} found on the mixture.

\subsection{Resource Accounting}\label{sec:resource}
Table~\ref{tab:resource} collects what each configuration costs per generated sample, so that the reductions claimed above can be read against the circuit that produces them. The gate counts follow from the $L$ layers of Fig.~\ref{fig:circuit}, each contributing $2n_q$ single-qubit rotations and a ring of $n_q$ CNOT gates. The baseline is the shot-noise-free readout, which is what a sampler pays when the finite-shot error is suppressed rather than absorbed. For the logarithmic encoding one execution of the circuit yields one shot, so the two rows coincide. For the per-dimension encoding the budget \eqref{M8q} is allocated per coordinate while a single execution reads every qubit at once, so the number of executions is set by the most demanding coordinate and slightly exceeds the mean shot count.

\begin{table}[t]
\centering
\caption{Per-sample resources of the four quantum configurations.}
\label{tab:resource}
\begin{tabular}{lcccc}
\toprule
 & \multicolumn{2}{c}{Gaussian mixture} & \multicolumn{2}{c}{MNIST latent} \\
\cmidrule(lr){2-3}\cmidrule(lr){4-5}
 & log. & per-dim & log. & per-dim \\
\midrule
qubits $n_q$                 & $3$   & $8$   & $3$   & $8$   \\
re-uploading layers $L$      & $12$  & $12$  & $20$  & $12$  \\
single-qubit gates           & $72$  & $192$ & $120$ & $192$ \\
two-qubit gates              & $36$  & $96$  & $60$  & $96$  \\
reverse steps                & $20$  & $20$  & $20$  & $20$  \\
shots per sample             & $303$ & $991$ & $430$ & $993$ \\
circuit executions           & $303$ & $1004$& $430$ & $1004$\\
noise-free baseline          & \multicolumn{4}{c}{$1.6\times10^{6}$} \\
reduction                    & $5300\times$ & $1600\times$ & $3700\times$ & $1600\times$ \\
\bottomrule
\end{tabular}
\end{table}

\section{Conclusion}
\gthchanged{We have shown that finite-shot quantum measurement noise can supply part of the randomness a stochastic diffusion sampler would otherwise inject. Three elements make this possible. A bounded zero-sum representation mapped to the probability simplex connects the circuit's Born probabilities to the denoising task. A trainable linear readout determines how measurement fluctuations enter the reverse update. A PSD condition then identifies how much of the prescribed noise covariance these fluctuations can supply, with classical noise filling the remainder. Under the stated assumptions, this compensation matches the conditional moments of the linear update; distribution matching additionally relies on a Gaussian approximation. Solving the condition for the shot count gives a state-dependent adaptive budget that increases as the sampler becomes more deterministic. On an eight-dimensional Gaussian mixture, the adaptive schedule achieves comparable generation quality with $303$ shots per sample against $1.6\times10^6$ for the high-shot reference. A uniform allocation needs about four times as many shots to match it, and the same mechanism supports MNIST generation in a learned latent space.}

\gthchanged{Several directions remain open. The framework assumes a linear readout, and whether a nonlinear one admits a comparable covariance contract is not obvious. The budget bounds are set by hand. Further analysis is needed to quantify the effects of non-Gaussian finite-shot errors, estimated residual covariances, boundary projections, and denoiser approximation on the generated distribution. Finally, the scheme has so far been validated in simulation, where finite-shot noise is the only source of error. On hardware it coexists with gate and decoherence errors, whose covariance is not multinomial, and whether the reverse process can absorb those as well is the natural next question.}

\printbibliography
\end{document}